\documentclass[twocolumn]{aastex631}
\usepackage{natbib}
\usepackage{amsmath}
\usepackage{gensymb}
\received{8 December 2022}
\revised{21 November 2023}
\accepted{27 November 2023}
\acceptjournal{Astronomical Journal}

\shorttitle{Multiple-planet Systems Near Mean-motion Resonances Are Young}
\shortauthors{Hamer \& Schlaufman}

\begin{document}

\title{Kepler-discovered Multiple-planet Systems Near Period Ratios
Suggestive of Mean-motion Resonances Are Young}

\correspondingauthor{Jacob H.\ Hamer}
\email{jacobhhamer@gmail.com}

\author[0000-0002-7993-4214]{Jacob H.\ Hamer}
\affiliation{New Jersey State Museum, 205 W State St, Trenton, NJ
08608, USA}
\affiliation{William H.\ Miller III Department of Physics \& Astronomy,
Johns Hopkins University, 3400 N Charles St, Baltimore, MD 21218, USA}

\author[0000-0001-5761-6779]{Kevin C.\ Schlaufman}
\affiliation{William H.\ Miller III Department of Physics \& Astronomy,
Johns Hopkins University, 3400 N Charles St, Baltimore, MD 21218, USA}

\begin{abstract}

\noindent
Before the launch of the Kepler Space Telescope, models of low-mass
planet formation predicted that convergent Type I migration would often
produce systems of low-mass planets in low-order mean-motion resonances.
Instead, Kepler discovered that systems of small planets frequently have
period ratios larger than those associated with mean-motion resonances and
rarely have period ratios smaller than those associated with mean-motion
resonances.  Both short-timescale processes related to the formation or
early evolution of planetary systems and long-timescale secular processes
have been proposed as explanations for these observations.  Using a thin
disk stellar population's Galactic velocity dispersion as a relative age
proxy, we find that Kepler-discovered multiple-planet systems with at
least one planet pair near a period ratio suggestive of a second-order
mean-motion resonance have a colder Galactic velocity dispersion and
are therefore younger than both single-transiting and multiple-planet
systems that lack planet pairs consistent with mean-motion resonances.
We argue that a non-tidal secular process with a characteristic timescale
no less than a few hundred Myr is responsible for moving systems of
low-mass planets away from second-order mean-motion resonances.  Among
systems with at least one planet pair near a period ratio suggestive
of a first-order mean-motion resonance, only the population of systems
likely affected by tidal dissipation inside their innermost planets has
a small Galactic velocity dispersion and is therefore young.  We predict
that period ratios suggestive of mean-motion resonances are more common
in young systems with $10\text{~Myr}\lesssim\tau\lesssim100\text{~Myr}$
and become less common as planetary systems age.

\end{abstract}

\keywords{Exoplanet dynamics(490) --- Exoplanet evolution(491) ---
Exoplanet systems(484) --- Exoplanet tides(497) ---
Exoplanets(498) --- Star-planet interactions(2177) ---
Stellar ages(1581) --- Stellar kinematics(1608) ---
Tidal interaction(1699)}

\section{Introduction}

Gravitationally mediated interactions between newly formed planets
and their parent protoplanetary disks have long been predicted to
cause planets to radially migrate within their parent protoplanetary
disks \citep[e.g.,][]{Goldreich1980,Ward1986,Ward1997,Lin1986}.
Further studies of these processes led to the prediction
that the trajectories of migrating planets would converge and
result in dynamically important planet--planet interactions
\citep[e.g.,][]{Bryden2000,Kley2000,Masset2001}.  The Doppler-based
discovery of the resonant multiple giant planet system GJ 876
\citep{Marcy2001} confirmed these predictions for convergent
migration and showed that they can lead to resonant configurations.
More detailed studies of the resonant capture process have since
suggested that capture into low-order mean-motion resonances is a likely
outcome if migration rates are slow and eccentricity damping efficient
\citep[e.g.,][]{Snellgrove2001,Lee2002,Nelson2002,Thommes2005,Kley2005,Quillen2006,Pierens2008,Lee2009,Ogihara2013}.
Planet--planet scattering rarely leads to mean-motion resonances
\citep[e.g.,][]{Raymond2008}.  Though there are no planet--planet
mean-motion resonances in the solar system\footnote{Jupiter and
Saturn are near a 5:2 mean-motion resonance but are not resonant
\citep[e.g.,][]{Michtchenko2001}.  There are mean-motion resonances in
giant planet satellite systems that can be explained by differential
tidal migration quite unlike the disk-mediated processes described above
\citep[e.g.,][]{Goldreich1965}.}, multiple giant exoplanet systems with
period ratios suggestive of mean-motion resonances are more common than
expected if orbital periods were uncorrelated \citep[e.g.,][]{Wright2011}.

The successful explanation of mean-motion resonances in multiple
giant planet systems by the convergent migration scenario motivated
its extension to systems of low-mass planets and the expectation that
mean-motion resonances would be relatively common in systems of low-mass
planets \citep[e.g.,][]{Papaloizou2005,Cresswell2006,Terquem2007}.
This turned out not to be the case.  Data from the Kepler Space Telescope
\citep{Borucki2010} revealed that while multiple small planet systems
are more likely to be found with period ratios suggestive of first-order
mean-motions resonances than if period ratios were randomly distributed,
period ratios suggestive of mean-motion resonances are not common
\citep[e.g.,][]{Lissauer2011,Fabrycky2014}.  Instead, planet pairs tend
to favor period ratios just wide of resonances and tend to avoid period
ratios just interior to resonances.

The properties of the Kepler-discovered multiple small planet
system period distribution have been attributed to short-timescale
processes related to the formation or early evolution of planetary
systems.  Multiple small planet systems that formed in situ would
have no obvious reason to be found with period ratios suggestive of
mean-motion resonances, though planet--planet interactions might
still result in mean-motion resonances during in situ formation
\citep[e.g.,][]{Petrovich2013}.  Mean-motion resonances initially
present in systems formed via convergent migration might be forced
out of resonance by turbulence in their parent protoplanetary
disks \citep[e.g.,][]{Adams2008,Rein2012,Batygin2017}
or by density waves generated by the planets themselves
\citep[e.g.,][]{PodlewskaGaca2012,Baruteau2013,Cui2021}.
The migration and eccentricity damping timescales present in real
protoplanetary disks may make resonant configurations transient
\citep[e.g.,][]{Goldreich2014,Charalambous2022,Laune2022}.  Dynamical
interactions with residual planetesimals can produce period ratios
wide of mean-motion resonances as well \citep[e.g.,][]{Thommes2008,
Chatterjee2015,Ghosh2023}.

Long-timescale secular processes have also been proposed to explain the
properties the Kepler-discovered multiple small planet system period
distribution, especially secular interactions between near-resonant
planets in the presence of weak dissipation.  Because tidal dissipation
can remove orbital energy from inner planets much more efficiently
than from outer planets, a planet pair subject to differential
tidal dissipation will naturally evolve to a larger period ratio
\citep[e.g.,][]{Papaloizou2011,Lithwick2012,Batygin2013b,Delisle2012,Delisle2014a,Lee2013}.
As expected in this tidal dissipation scenario, \citet{Delisle2014b}
found that as the period of the innermost planet in a Kepler-discovered
system of small planets increases, the excess of planet pairs exterior
to period ratios suggestive of mean-motion resonances decreases.
In contrast, \citet{Lee2013} showed that tidal dissipation must be an
order of magnitude more efficient than expected to explain the properties
of the Kepler-discovered multiple small planet period distribution.
\citet{Silburt2015} further argued that in the tidal dissipation scenario
the initial eccentricities required to explain the occurrence of planets
exterior to 2:1 period ratios are unreasonably high.

A comparison of the relative ages of Kepler-discovered small planet
systems close to/far from period ratios suggestive of mean-motion
resonances would help differentiate between these possibilities.
Since most Kepler-discovered systems of small planets orbit mature stars
with ages $\tau \gtrsim 1$ Gyr, the lack of an age offset between systems
with planets close to/far from period ratios suggestive of mean-motion
resonances would imply that systems move out of mean-motion resonances
early in their histories at ages $\tau \lesssim 1$ Gyr.  On the other
hand, the observation that systems with planets far from period ratios
suggestive of mean-motion resonances are older than those systems with
planets close to period ratios suggestive of mean-motion resonances would
support the idea that long-timescale secular processes move systems out
of mean-motion resonance.  Additional secular processes beyond tidal
dissipation would be necessary if systems with both small-separation
and large-separation innermost planets displayed the same relationship
between relative age and separation from period ratios suggestive of
mean-motion resonances.

In this article, we use the Galactic velocity dispersion of a thin disk
stellar population as a proxy for its age and find that Kepler-discovered
systems of small planets with period ratios suggestive of second-order
mean-motion resonances are younger than both single-transiting and
multiple-planet systems that lack planets with period ratios suggestive
of mean-motion resonances.  We suggest that non-tidal secular processes
operating over hundreds of Myr to Gyr are responsible for driving planet
pairs away from period ratios suggestive of second-order mean-motion
resonances.  Among systems with at least one planet pair near a period
ratio suggestive of a first-order mean-motion resonance, only the
population of systems likely affected by tidal dissipation inside
their innermost planets have a small Galactic velocity dispersion
and are therefore young.  We describe in Section 2 the assembly of
our analysis sample.  We detail in Section 3 the methodology we use to
identify systems with period ratios suggestive of mean-motion resonances,
evaluate the importance of tides for a system's evolution, and compare
the relative ages of systems close to/far from period ratios suggestive of
mean-motion resonances.  We review the explanations for and implications
of our analyses in Section 4.  We conclude by summarizing our findings
in Section 5.

\section{Data}

We use in our analysis a catalog of Kepler planet candidates optimized
for completeness as well as the detection and characterization
of planet candidates in the presence of transit-timing variations
\citep{Lissauer2023}.\footnote{We have verified that our conclusions would
be unchanged if we instead used the \citet{Thompson2018} Kepler Data
Release 25 (DR25) Kepler Objects of Interest (KOI) list optimized for
uniformity and statistical analyses and rejected all planet candidates
classified as false positives.}  Our sample includes all confirmed and
candidate planets with a disposition `P' in the \citet{Lissauer2023}
catalog.  Because virtually all Kepler-discovered planet candidates in
multiple-planet systems are true planets \citep[e.g.,][]{Lissauer2012},
we analyze all multiple-planet systems regardless of whether their
constituent planets have been classified as planets or planet candidates.

Following \citet{Schlaufman2021}, we calculate updated planet radii
using the observed transit depths in our input catalog and the
stellar radii derived in \citet{Brewer2018}, \citet{Johnson2017},
or \citet{Berger2020} (in order of decreasing priority).  We account
for both transit depth and stellar radius uncertainties in our updated
planet radius uncertainties.  We collect Gaia DR3 \texttt{source\_id}
information for our sample using \texttt{astroquery} \citep{Ginsburg2019}
to query SIMBAD \citep{Wenger2000} for each Kepler Input Catalog (KIC)
identifier in our sample.  We use each Gaia DR3 \texttt{source\_id}
to collect all other relevant information for each star in our
catalog from the Gaia Archive\footnote{For the details of Gaia DR3
and its data processing, see \citet{Gaia0,GaiaE3,Gaia3,Gaia3frame},
\citet{GaiaE3validation}, \citet{GaiaE3bias,GaiaE3astrometric},
\citet{GaiaE3xmatch,Gaia3xmatch}, \citet{GaiaE3photometry},
\citet{GaiaE3psf}, \citet{GaiaE3source}, \citet{Gaia3cat}, and
\citet{Gaia3rv}.} and require $\texttt{parallax\_over\_error} > 10$
to ensure the precision of our velocity dispersion inferences.

In addition to Gaia astrometry, we need radial velocities to
calculate the space velocities of individual stars and thereby the
velocity dispersion of our sample.  In order of decreasing priority,
we use the radial velocities from the California-Kepler-Survey
\citep[CKS;][]{Petigura2017}, the Apache Point Observatory Galactic
Evolution Experiment (APOGEE)\footnote{Based on spectra that
were gathered during the third and fourth phases of the Sloan
Digital Sky Survey \citep[SDSS;][]{Eisenstein2011,Blanton2017}
as part of the APOGEE effort \citep{Majewski2017}.
These spectra were collected with the APOGEE spectrographs
\citep{Zasowski2013,Zasowski2017,Pinsonneault2014,Wilson2019,Beaton2021,Santana2021}
on the New Mexico State University 1-m Telescope \citep{Holtzman2010}
and the Sloan Foundation 2.5-m Telescope \citep{Gunn2006}.  As part of
SDSS DR17 \citep{APOGEEDR17}, these spectra were reduced and analyzed
with the APOGEE Stellar Parameter and Chemical Abundance Pipeline
\citep[ASPCAP;][]{Holtzman2015,Nidever2015,Garcia2016}.}, DR7 of the
Large Sky Area Multi-Object Fiber Spectroscopic Telescope (LAMOST)
Medium-Resolution Spectroscopic Survey \citep[MRS;][]{LAMOST,LAMOST2},
Gaia DR3 \citep{Gaia3rv}, and DR7 of the LAMOST Low-Resolution
Spectroscopic Survey \citep[LRS;][]{LAMOST, LAMOST2}.  We combine multiple
APOGEE radial velocities in the table \texttt{apogeeStar} for a single
object using a \texttt{vscatter}-weighted average.  We combine multiple
LAMOST radial velocities for a single object using a weighted average,
weighting by radial velocity uncertainty for the MRS and by the $g$-band
spectral signal-to-noise ratio for the LRS.  If the difference between the
maximum and minimum LAMOST radial velocities for a single object is larger
than three times the uncertainty of the weighted average, we do not use
the LAMOST radial velocity.  Following \citet{Marchetti2022}, to ensure
reliable Gaia DR3 radial velocities we require $\texttt{rv\_nb\_transits}
> 10$ and $\texttt{rv\_expected\_sig\_to\_noise} > 5$.

The presence of short-period stellar-mass binaries in a comparison
sample unlikely to be present in a sample of transiting multiple-planet
systems can artificially inflate the velocity dispersion of the
comparison sample.  We therefore eliminate from our analysis
sample all primary stars with radial velocity variations best
explained by the Keplerian orbits of stellar-mass companions
with unimodal posteriors in the APOGEE-based catalog presented in
\citet{PriceWhelan2017,Price-Whelan2020}\footnote{\url{https://www.sdss.org/dr17/data_access/value-added-catalogs/?vac_id=orbital-parameter-samplings-of-apogee-2-stars-from-the-joker}}.
We thereby identify two stars in our sample which may have binary
companions: Kepler-470/KOI-129/KIC 11974540 and Kepler-1717/KOI-353/KIC
11566064.  The system Kepler-470 has maximum-a-posteriori (MAP)
Doppler-inferred orbital period $P_{\text{Dop}} = 24.70$ days, very
close to the transit-inferred orbital period of Kepler-470 b/KOI-129.01
$P_{\text{tra}} = 24.67$ days.  Assuming the star Kepler-470 has a
stellar mass $M_{\ast} = 1.33~M_{\odot}$, the system's MAP period,
Doppler semiamplitude $K = 5.79$ km s$^{-1}$, and eccentricity $e =
0.05$ imply that the transiting object Kepler-470 b has a mass $M
\gtrsim 0.1~M_{\odot}$.  We therefore remove it from our sample of
planet-mass objects.  The system Kepler-1717/KOI-353/KIC 11566064 has a
MAP Doppler-inferred orbital period $P_{\text{Dop}} = 40.72$ days that
is incompatible with the transit-inferred orbital periods of KOI-353.01,
Kepler-1717 b/KOI-353.02, or KOI-353.03 at  $P_{\text{tra}} =$ 11.16,
30.65, and 152.10 days.  Assuming the star Kepler-470/KIC 11974540 has a
stellar mass $M_{\ast} = 1.35~M_{\odot}$, the system's MAP orbital period,
Doppler semiamplitude $K = 1.82$ km s$^{-1}$, and eccentricity $e = 0.87$
imply the presence of a secondary with $M \gtrsim 20~M_{\text{Jup}}$.
The presence in the system of an object with these Keplerian orbital
parameters---possibly a non-transiting, highly eccentric brown dwarf---is
difficult to reconcile the properties of the confirmed planet Kepler-1717
b.  In this case we decided that the transit-based system parameters
are likely more realistic and therefore keep the system in our sample.

While the vast majority of known planet host stars are members of
the Milky Way's thin disk stellar population, the presence in our
comparison or analysis samples of planet host stars belonging to the
thick disk or halo stellar populations could significantly inflate their
velocity dispersions.  We identify the star Kepler-292/KOI-1364/KIC
6962977 as a kinematic member of the thick disk after using
\texttt{galpy}\footnote{\url{http://github.com/jobovy/galpy}}
\citep{Bovy2015} to integrate its orbit in the \citet{McMillan2017} Milky
Way potential over several Gyr.  We find that Kepler-292/KOI-1364/KIC
6962977 has $z_{\text{max}} \gtrsim 3$ kpc and $e = 0.47$ and therefore
remove it from our sample.  Our final analysis sample of Kepler-discovered
transiting multiple-planet systems with the necessary data to calculate
precise Galactic kinematics includes 576 systems with 1460 planets.

\section{Analysis}

\subsection{Identification of Systems with a Plausibly-resonant Planet
Pair}

It is non-trivial to prove that a pair of planets with a period ratio
suggestive of a mean-motion resonance is actually resonant with librating
orbital elements.  Detailed dynamical analyses have shown that planet
pairs with period ratios close to but not exactly equal to rational
numbers may indeed be resonant with librating orbital elements.  This is
the case for GJ 876 \citep[e.g.,][]{Peale1976, Rivera2001}.  Similarly,
planet pairs with period ratios apparently equal to rational numbers may
in reality have circulating orbital elements \citep[e.g.,][]{Veras2012}.
In short, detailed dynamical modeling is required to prove that a pair
of planets is resonant with librating orbital elements.  These dynamical
models critically depend on planet masses and orbital eccentricities
that are often unknown or poorly constrained for low-mass planets
that represent the majority of known exoplanet systems.  As a result,
approximate methods are often necessary to evaluate whether or not a
pair of low-mass planets can be thought of as plausibly resonant.

Because Kepler-discovered systems of small planets rarely have the
precise constraints on planet masses and orbital elements necessary
for rigorous dynamical modeling, heuristic methods have often been
used to identify pairs of planets plausibly in mean-motion resonances.
Virtually all heuristic methods in the literature assume that planet
pairs with period ratios closest to rational numbers are also most
likely to be in mean-motion resonances.  We use two different heuristic
methods to identify plausibly resonant pairs of planets, one that is more
physically motivated but that depends on mostly unknown planet masses
and orbital eccentricities and one that is less physically motivated but
independent of planet masses and orbital eccentricity.  As we will show,
we reach the same conclusions using either methodology.

\begin{figure*}
\includegraphics[width=\linewidth]{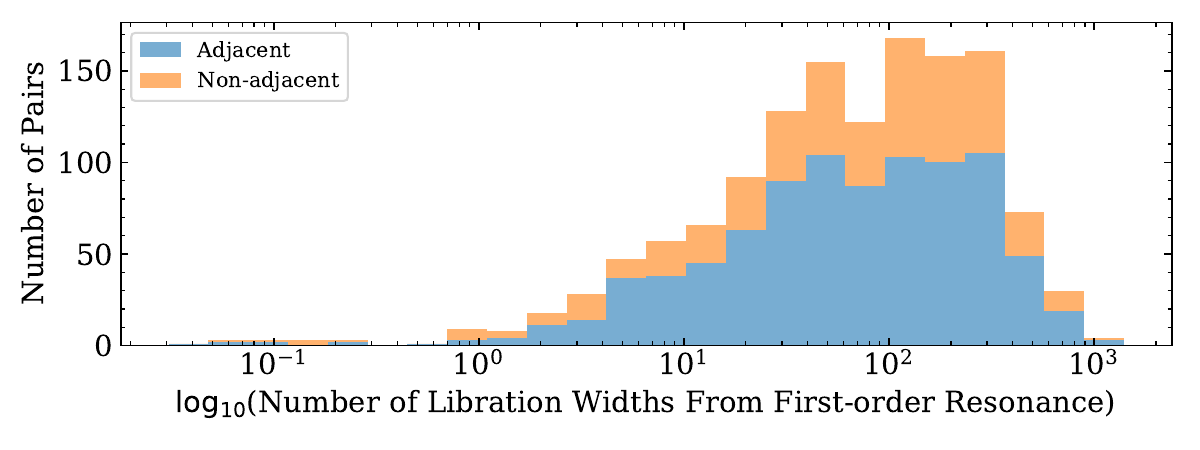}
\includegraphics[width=\linewidth]{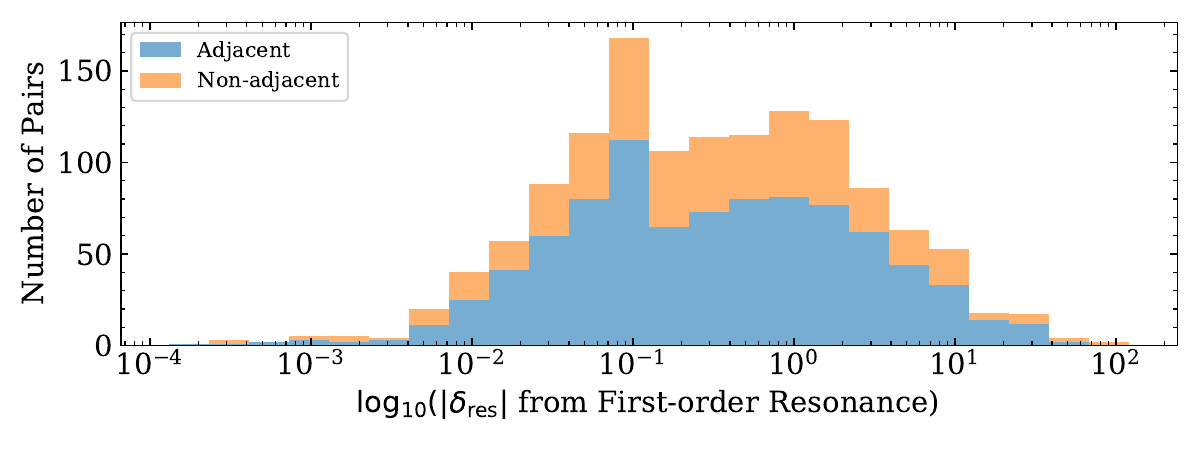}
\includegraphics[width=\linewidth]{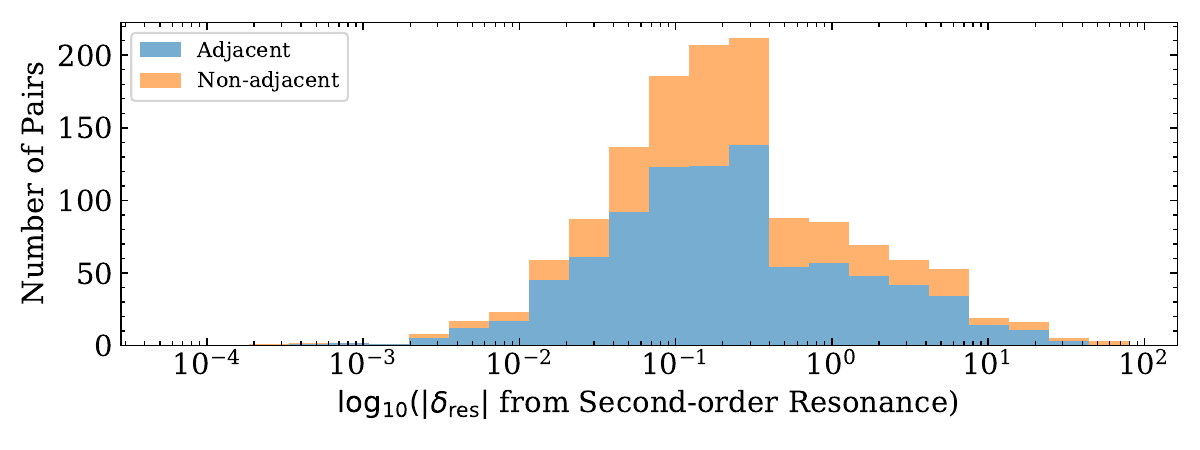}
\caption{Distributions of separation from closest mean-motion resonance.
In each panel the blue histogram gives the number of adjacent planet
pairs in each histogram bin, while the orange histogram stacked on top
of the blue histogram gives the number of non-adjacent planet pairs in
each histogram bin.  Top: distribution of separation from first-order
mean-motion resonances in units of libration widths.  We consider planet
pairs within five resonance libration widths ``plausibly resonant''
in our subsequent analyses.  Middle: distribution of separation from
first-order mean-motion resonances in $\delta_{\text{res}}$.  Bottom:
distribution of separation from second-order mean-motion resonances in
$\delta_{\text{res}}$.  We consider planet pairs with $\delta_{\text{res}}
< 0.015$ plausibly resonant in our subsequent analyses.  Since our
definition of plausibly resonant refers only to mean-motion resonances,
from here the phrase plausibly resonant implies plausibly mean-motion
resonant.\label{figure01}}
\end{figure*}

The first method we use to identify plausibly resonant planet pairs is the
physically-motivated but planet mass- and orbital eccentricity-dependent
heuristic method proposed by \citet{Steffen2015}.  The \citet{Steffen2015}
approach considers planet pairs with period ratios separated from
first-order mean-motion resonances by a small number of libration widths
plausibly resonant, while planet pairs with period ratios separated
from first-order mean-motion resonances by a large number of libration
widths non-resonant.  For each planet pair we calculate libration widths
normalized to semimajor axis using Equation (8.76) of \citet{Murray1999}
\begin{eqnarray}
\frac{\delta a_\text{max}}{a} & = &
	\pm \left(\frac{16}{3}\frac{|C_r|}{n}e\right)^{1/2}
	\left(1 + \frac{1}{27j_2^2e^3}\frac{|C_r|}{n}\right)^{1/2} \\
   & & -\frac{2}{9j_2e}\frac{|C_r|}{n}\nonumber.
\end{eqnarray}
The ratio $C_r/n$ of the resonant part of the disturbing function $C_r$
to the orbital frequency $n$ is defined as $(m'/m_c) \alpha f_d(\alpha)$,
where $m_c$ is the mass of the host star, $m'$ is the mass of the
outer planet in a planet pair, and $\alpha f_d(\alpha)$ are numerical
factors from Table 8.5 of \citet{Murray1999}.  The quantity $j_{2} =
-p$ where $p$ is the denominator of the period ratio (e.g., 1 for a 2:1
mean-motion resonance).  We use the \citet{Lissauer2011} mass--radius
relation $M_{\text{p}} = (R_{\text{p}}/R_{\oplus})^{2.06}$ to infer
masses for each planet in a pair based on their radii and calculate
the value of the inner planet's eccentricity that minimizes libration
width.  This latter assumption ensures that we are as conservative as
possible with our inferred libration widths.  We plot in the top panel
of Figure \ref{figure01} the distribution of separations from the closest
first-order mean-motion resonance in units of libration widths.  We note
that the libration width relation was derived under the assumption of a
massive outer planet and a massless inner planet.  While this assumption
is not strictly valid, \citet{Steffen2015} reasoned that since libration
width generally scales with total planet mass \citep[e.g.,][]{Deck2013}
this approach is still valid.\footnote{While not directly applicable to
our problem, more complete theoretical models accounting for non-zero
masses in both inner and out planets in a first-order mean-motion resonant
pair can be found in \citet{Batygin2013c} and \citet{Nesvorny2016}.}

The \citet{Steffen2015} heuristic method for the identification of
plausibly resonant planet pairs described above is planet mass- and
orbital eccentricity-dependent.  An alternative planet mass- and orbital
eccentricity-independent heuristic is the $\epsilon$ parameter defined
for first- and second-order resonances as
\begin{eqnarray}
\epsilon_{1} & = & \frac{P_{\text{out}}}{P_{\text{in}}} - \frac{j+1}{j}, \\
\epsilon_{2} & = & \frac{P_{\text{out}}}{P_{\text{in}}} - \frac{j+2}{j},
\end{eqnarray}
where $P_{\text{out}}/P_{\text{in}}$ is the ratio of the outer planet
period to the inner planet period.  The $\epsilon$ parameter is
often used to describe the dynamics of systems near mean-motion
resonances, and predictions for the evolution of $\epsilon$
away from resonances in initially resonant systems as a result of
tidal dissipation or planetesimal scattering have been published
\citep{Delisle2014b,Chatterjee2015}.  The $\epsilon$ parameter does not
account for the varying widths of resonances though, so a fixed separation
in $\epsilon$ could be far from a relatively narrow 6:5 resonance but
close to a relatively wide 2:1 resonance.  In contrast, the $\zeta$
parameter described in \citet{Lissauer2011} and \citet{Fabrycky2014}
defined for first- and second-order resonances as
\begin{eqnarray}
\zeta_{1} & = & 3\left[\frac{1}{P_{\text{out}}/P_{\text{in}}-1}-\text{Round}\left(\frac{1}{P_{\text{out}}/P_{\text{in}}-1}\right)\right], \\
\zeta_{2} & = & 3\left[\frac{2}{P_{\text{out}}/P_{\text{in}}-1}-\text{Round}\left(\frac{2}{P_{\text{out}}/P_{\text{in}}-1}\right)\right],
\end{eqnarray}
accounts for the varying widths of first- and second-order resonances
and is able to highlight asymmetries in the period ratio distribution of
Kepler-discovered systems of small planets near first-order resonances.
There is no physically-motivated way to specify the probability that a
planet pair is truly in resonance as a function of $\zeta$ though.

The second method we use to identify plausibly resonant planet pairs is
an extension of the $\epsilon$ parameter-based method described above.
We define the parameter
\begin{eqnarray}
\delta_{\text{res}} = \frac{1}{p_{\text{res}}}
	              \left(\frac{P_{\text{out}}}{P_{\text{in}}}-
		      p_{\text{res}}\right),
\end{eqnarray}
where $p_{\text{res}} = (j + i)/j$ and $i \in (1,2)$.  This
$\delta_{\text{res}}$ parameter differs from the $\epsilon$ parameter
used in \citet{Delisle2014b} and \citet{Chatterjee2015} in that it is
normalized by the resonant period ratio.  While this normalization has
been neglected in theoretical analyses of departures from resonances
\citep[e.g.,][]{Lee2013}, we include this normalization to account for
the fact that widely-spaced resonances have larger ``widths'' in period
ratio than more closely-spaced resonances.  For example, a 3\% fractional
period-ratio separation from a 2:1 resonance would extend over 0.06 in
period ratio while the same fractional period-ratio separation from a
6:5 resonance would extend over only 0.036 in period ratio.  We plot
in the bottom two panels of Figure \ref{figure01} the distributions of
separation from the closest first- or second-order mean-motion resonance
in terms of $\delta_{\text{res}}$.

For the libration width-based heuristic we classify as ``plausibly
resonant'' planet pairs within five libration widths of a single
mean-motion  resonance.  Because the $\delta_{\text{res}}$-based
heuristic lacks a physically motivated threshold for classification
as plausibly resonant, we classify as plausibly resonant planet pairs
with $\delta_{\text{res}} < 0.015$ to roughly match the size of the
libration width-based plausibly mean-motion resonant sample.  We fully
acknowledge that the planet pairs pairs we classify as plausibly resonant
may not have librating orbital elements and therefore may not truly be
in mean-motion resonances.  Since our definition of plausibly resonant
refers only to mean-motion resonances, from here the phrase plausibly
resonant implies plausibly mean-motion resonant.

Because it is well-established that
resonance overlap leads to chaotic behavior
\citep[e.g.,][]{Wisdom1980,Lecar2001,Quillen2011,Deck2013,Batygin2013b,Morrison2016},
if a pair of planets is within five libration widths or
$\delta_{\text{res}} < 0.015$ of multiple resonances then we do not
consider the pair to be plausibly resonant.  As the widths of resonances
increase with pair total mass, the existence of planet pairs apparently
occupying multiple resonances may imply that the mass--radius relation we
used for the libration width-based definition overestimated some planet
pairs' total masses.  In support of this possibility, we find that
planet pairs occupying multiple resonances according to the libration
width-based definition are more massive than plausibly resonant pairs or
pairs outside of resonance entirely.  These planet pairs apparently in
multiple resonances likely instead have lower masses than predicted by
the \citet{Lissauer2011} mass--radius relation, indicating the existence
of a large number of planets with $2~R_{\oplus} \lesssim R_{\text{p}}
\lesssim 4~R_{\oplus}$ that are less dense than Uranus and Neptune in
our own solar system \citep[e.g.,][]{Schlaufman2021}.  We also remove
from our sample of plausibly resonant systems two-planet systems that
dynamical analyses by \citet{Veras2012} have shown must be circulating
instead of librating under the assumption that the two observed planets
are the only planets in the system.

\begin{figure*}
\includegraphics[width=\linewidth]{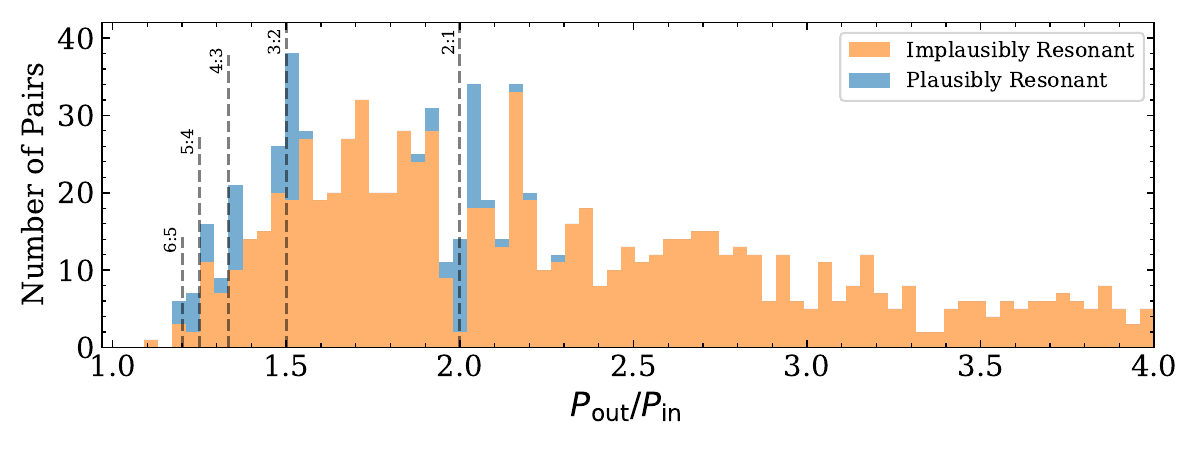} \\
\includegraphics[width=\linewidth]{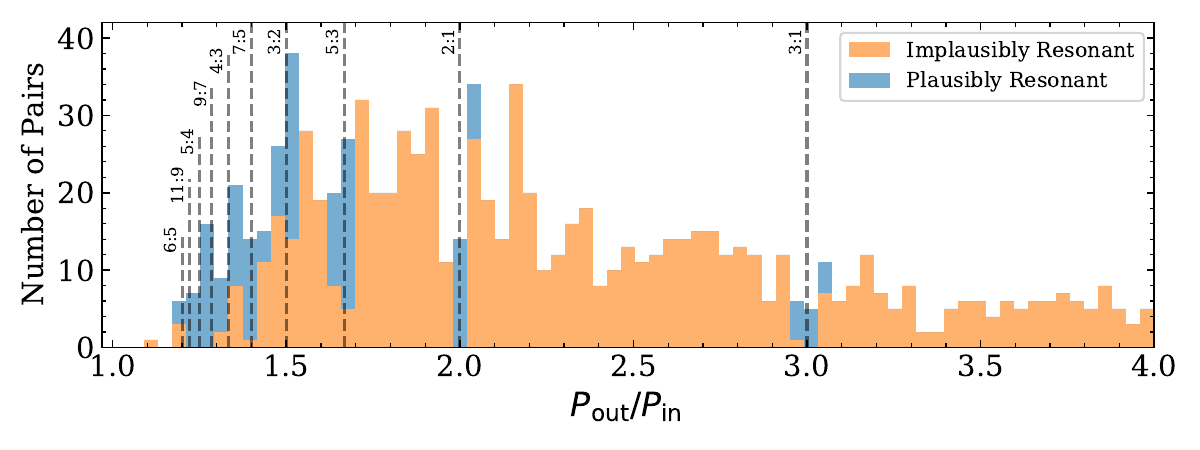}
\caption{Distribution of planet-pair period ratios.  The orange histogram
gives the number of implausibly resonant pairs in each period ratio
bin while the blue histogram stacked on top of the orange histogram
gives the number of plausibly resonant pairs in each period ratio bin.
Top: plausibly resonant/implausibly resonant according to the libration
width-based definition.  Plausibly resonant pairs far from period ratios
suggestive of resonances are high total-mass planet pairs with larger
libration widths.  Bottom: plausibly resonant/implausibly resonant
according to the $\delta_{\text{res}}$-based definition.\label{figure02}}.
\end{figure*}

For each planet pair in our sample, we report in Table \ref{table01}
separations in units of libration widths from each first-order resonance
we considered.  In Table \ref{table02} we report the separations from
the same first-order and additional second-order resonances in terms
of $|\delta_{\text{res}}|$.  We list in Table \ref{table03} the number
of plausibly resonant pairs in each first- and second-order resonance
we considered.  We plot in Figure \ref{figure02} stacked histograms
comparing the period ratio distributions of our plausibly resonant and
implausibly resonant samples.

\begin{deluxetable*}{cccccccccc}
\tablecaption{Separations from Resonances in Libration
Widths\label{table01}}
\tablehead{
\colhead{KIC} &
\colhead{Inner KOI} &
\colhead{Outer KOI} &
\colhead{Adjacent} &
\colhead{2:1} &
\colhead{3:2} &
\colhead{4:3} &
\colhead{5:4} &
\colhead{6:5} &
\colhead{Plausibly Resonant}}
\tablenum{1}
\startdata
 1432789 &   992.01 &   992.02 &   True &  20 &  77 &  92 &  97 &  98  & False\\
 1717722 &  3145.01 &  3145.02 &   True & 311 & 361 & 361 & 352 & 342  & False\\
 1718189 &   993.03 &   993.01 &   True & 369 & 452 & 459 & 451 & 439  & False\\
 1718189 &   993.03 &   993.02 &  False & 554 & 596 & 584 & 564 & 544  & False\\
 1718189 &   993.01 &   993.02 &   True &  96 &  50 &  94 & 112 & 120  & False\\
 1724719 &  4212.02 &  4212.01 &   True & 112 & 105 & 169 & 194 & 205  & False\\
 1871056 &  1001.02 &  1001.01 &   True & 143 & 189 & 196 & 194 & 190  & False\\
 1996180 &  2534.02 &  2534.01 &   True & 118 & 164 & 245 & 277 & 289  & False\\
 2165002 &   999.02 &   999.01 &   True &  93 & 144 & 154 & 155 & 153  & False\\
 2302548 &   988.02 &   988.01 &   True &  52 & 122 & 139 & 143 & 143  & False
\enddata
\tablecomments{The number of libration widths from each first-order
resonance considered for each planet pair in the sample ordered by Kepler
Input Catalog (KIC) identifier.  We also provide indicators whether
or a not planet pair is adjacent or plausibly resonant according to
our libration-width based definition.  This table is published in its
entirety in the machine-readable format.  A portion is shown here for
guidance regarding its form and content.}
\end{deluxetable*}

\begin{splitdeluxetable*}{ccccccccBcccccccc}
\tablecaption{Separations from Resonance in $|\delta_\text{res}|$
\label{table02}}
\tablehead{
\colhead{KIC} & 
\colhead{Inner KOI} & 
\colhead{Outer KOI} &
\colhead{Adjacent} & 
\colhead{2:1} & 
\colhead{3:2} & 
\colhead{4:3} &
\colhead{5:4} & 
\colhead{6:5} & 
\colhead{3:1} & 
\colhead{5:3} & 
\colhead{7:5} & 
\colhead{9:7} & 
\colhead{11:9} &
\colhead{First-order} &
\colhead{Second-order} \\
\colhead{} & 
\colhead{} & 
\colhead{} &
\colhead{} & 
\colhead{} & 
\colhead{} & 
\colhead{} &
\colhead{} & 
\colhead{} & 
\colhead{} & 
\colhead{} & 
\colhead{} & 
\colhead{} & 
\colhead{} &
\colhead{Resonant} &
\colhead{Resonant}}
\tablenum{2}
\startdata
1432789 &   992.01 &   992.02 &   True & 0.085 & 0.446 & 0.627 & 0.735 & 0.808 & 0.277 & 0.302 & 0.549 & 0.687 & 0.775 & False & False \\
 1717722 &  3145.01 &  3145.02 &   True & 1.321 & 2.095 & 2.482 & 2.714 & 2.869 & 0.547 & 1.785 & 2.316 & 2.611 & 2.798  & False & False \\
 1718189 &   993.03 &   993.02 &  False & 2.320 & 3.427 & 3.980 & 4.312 & 4.533 & 1.213 & 2.984 & 3.743 & 4.164 & 4.433  & False & False \\
 1718189 &   993.03 &   993.01 &   True & 0.984 & 1.646 & 1.976 & 2.175 & 2.307 & 0.323 & 1.381 & 1.835 & 2.086 & 2.247  & False & False \\
 1718189 &   993.01 &   993.02 &   True & 0.163 & 0.115 & 0.255 & 0.339 & 0.394 & 0.442 & 0.004 & 0.195 & 0.301 & 0.369  & False & True \\
 1724719 &  4212.02 &  4212.01 &   True & 0.128 & 0.162 & 0.308 & 0.395 & 0.453 & 0.419 & 0.046 & 0.245 & 0.356 & 0.427  & False & False \\
 1871056 &  1001.02 &  1001.01 &   True & 0.717 & 1.289 & 1.575 & 1.747 & 1.861 & 0.144 & 1.060 & 1.452 & 1.670 & 1.809  & False & False \\
 1996180 &  2534.02 &  2534.01 &   True & 0.104 & 0.195 & 0.344 & 0.434 & 0.493 & 0.403 & 0.075 & 0.280 & 0.394 & 0.466  & False & False \\
 2165002 &   999.02 &   999.01 &   True & 0.428 & 0.905 & 1.143 & 1.285 & 1.381 & 0.048 & 0.714 & 1.041 & 1.222 & 1.337  & False & False \\
 2302548 &   988.02 &   988.01 &   True & 0.183 & 0.578 & 0.775 & 0.893 & 0.972 & 0.211 & 0.420 & 0.691 & 0.841 & 0.937 & False & False \\
\enddata
\tablecomments{The absolute value of distance in $\delta_{\text{res}}$
from each first- and second-order resonance considered for each planet
pair in the sample ordered by KIC identifier.  We also provide indicators
whether or a not planet pair is adjacent or plausibly first-/second-order
resonant according to our $\delta_{\text{res}}$-based based definition.
This table is published in its entirety in the machine-readable format.
A portion is shown here for guidance regarding its form and content.}
\end{splitdeluxetable*}

\begin{deluxetable*}{ccc}
\tablecaption{Resonance-occupation Distribution\label{table03}}
\tablehead{\colhead{Resonance} &
\colhead{Number of Plausibly Resonant Pairs}&
\colhead{Number of Plausibly Resonant Pairs}\\&
\colhead{By Libration Width Criteria}&
\colhead{By $\delta_\text{res}$ Criteria}}
\tablenum{3}
\startdata
\phm{TestTestTe}3:1\phm{TestTestTe} & \phm{TestTestTe}$\cdots$\phm{TestTestTe} & \phm{TestTestTe}14\phm{TestTestTe}\\
2:1  & 35 & 21 \\
5:3  & $\cdots$ & 34 \\
3:2  & 23 & 33 \\
7:5  & $\cdots$ & 10 \\
4:3  &  4 &  0 \\
9:7  & $\cdots$ &  0 \\
5:4  &  5 &  0 \\
11:9 & $\cdots$ &  0 \\
6:5  &  1 &  0
\enddata
\tablecomments{The number of plausibly resonant pairs in the
first- and second-order mean-motion resonances we considered in
this analysis.  Resonances are ordered by decreasing period ratio
$P_\text{out}/P_\text{in}$, so planets with initial period ratios larger
than 3:1 experiencing convergent disk-driven migration should encounter
these resonances from top to bottom.}
\end{deluxetable*}

\subsection{Identification of Systems Plausibly Affected by Tidal
Dissipation}

Tidal dissipation is likely important for many of the Kepler-discovered
systems of small planets that comprise the majority of both our
plausibly resonant and implausibly resonant samples described in the
previous subsection.  To identify systems where tidal dissipation likely
plays a significant role in the evolution of systems' period ratios,
we calculate circularization times for the innermost planet in each
system in our sample.  The timescale for orbit circularization due to
tidal dissipation inside a planet can be written
\begin{eqnarray}
\tau_{e} = \frac{e}{\dot{e}} = \frac{1}{21\pi}
\frac{Q_\text{p}}{k_{2}} \frac{M_\text{p}}{M_{\ast}}
\left(\frac{a}{R_\text{p}}\right)^5 P,
\end{eqnarray}
where $Q_\text{p}$, $k_{2}$, $M_{\text{p}}$, $M_{\ast}$, $a$,
$R_\text{p}$, and $P$ are the planet tidal quality factor, planet Love
number, planet mass, stellar mass, semimajor axis, planet radius, and
orbital period \citep[e.g.,][]{Lee2013}.  Both $Q_\text{p}$ and $k_{2}$
are sensitively dependent on a planet's interior structure.  While useful
constraints on $Q_\text{p}$ and $k_{2}$ are available for solar system
planets, the internal structures of most exoplanets are poorly constrained
leaving exoplanet $Q_\text{p}$ and $k_{2}$ highly uncertain.

A solar metallicity star's luminosity $L_{\ast}$ scales like
\begin{eqnarray}
\frac{L_{\ast}}{L_{\odot}} = \left(\frac{M_\ast}{M_\odot}\right)^\alpha,
\end{eqnarray}
where $\alpha \approx 2.4$ for $M_{\ast} \lesssim 0.6~M_{\odot}$
and $\alpha \approx 4$ for $0.6~M_{\odot} \lesssim M_{\ast} \lesssim
6~M_{\odot}$ \citep[e.g.,][]{Lamers2017}.  The instellation incident on
an exoplanet $F_{\text{p}}$ in units of Earth insolations $F_{\oplus}$
is therefore
\begin{eqnarray}
\frac{F_{\text{p}}}{F_{\oplus}} = \frac{L_{\ast}}{L_{\odot}}
	                          \left(\frac{a}{1~\text{AU}}\right)^{-2}.
\end{eqnarray}
In our sample, the median innermost planet instellation is approximately
200 $F_{\oplus}$.  More than 90\% of the innermost planets in our sample
experience $F_{\text{p}} \gtrsim 30~F_\oplus$.  \citet{Fulton2018}
showed that planets with $R_\text{p} \lesssim 1.7~R_{\oplus}$
and $F_{\text{p}} \gtrsim 30~F_\oplus$ are likely rocky, as any
gaseous envelopes initially present would have been stripped by
photoevaporation \citep[e.g.,][]{Owen2017}, core-powered mass loss
\citep[e.g.,][]{Ginzburg2018}, or some other process.  In our sample,
innermost planets with $R_{\text{p}} \lesssim 1.7~R_\oplus$ are likely
rocky while innermost planets with $R_{\text{p}} \gtrsim 2.0~R_\oplus$
likely have at least some hydrogen/helium (H/He) atmospheres.

To evaluate the circularization timescale $\tau_{e}$ for innermost
planets in our sample with $R_{\text{p}} < 4~R_{\oplus}$, we assume
the dissipative properties of Mars.  We favor Mars over Venus and
Earth because its dissipative properties are both well constrained
and unaffected by the presence of oceans or a massive atmosphere
\citep[e.g.,][]{Tyler2021,Farhat2022}.  Based on detailed observations of
the orbits of Phobos and Deimos, \citet{Duxbury1982} found $30 \lesssim
Q_{\text{Mars}} \lesssim 130$.  We use the smaller value to
 ensure that we have identified all systems with innermost
planets plausibly affected by tidal evolution.  We confirm below the
these tidal parameter values do not qualitatively affect our results.
\citet{Yoder1995} suggested $k_{2} = 0.14$ for Mars.  While planets with
$R_{\text{p}} > 1.7~R_{\oplus}$ likely have H/He atmospheres, there are no
planets with similar structures in the solar system and their dissipative
properties are consequently poorly constrained.  We therefore use Mars's
properties for all innermost planets with $R_{\text{p}} < 4~R_{\oplus}$
that likely have a small fraction of their masses in H/He atmospheres.

To evaluate the circularization timescale for innermost planets in
our sample with $R_{\text{p}} \geq 4~R_{\oplus}$, we assume the
dissipative properties of Neptune.  Among the solar system's gas
and ice giant planets, Neptune is the most dissipative by at least
an order of magnitude.  The dynamics of Neptunian satellites suggest
$9000 \lesssim Q_\text{Neptune} \lesssim 36000$ assuming a theoretically
derived Neptunian $k_{2} = 0.41$ \citep{Bursa1992,Zhang2008}.  As before,
we adopt a value at the low end of the range.  While the assumption of
similar dissipative properties for exoplanets with $R_{\text{p}} \geq
4~R_{\oplus}$ might overestimate the important of tidal dissipation and
lead to underestimated circularization times, we use Neptune's dissipative
properties to ensure that we have identified all systems with innermost
planets plausibly affected by tidal evolution.

\begin{figure*}
\plotone{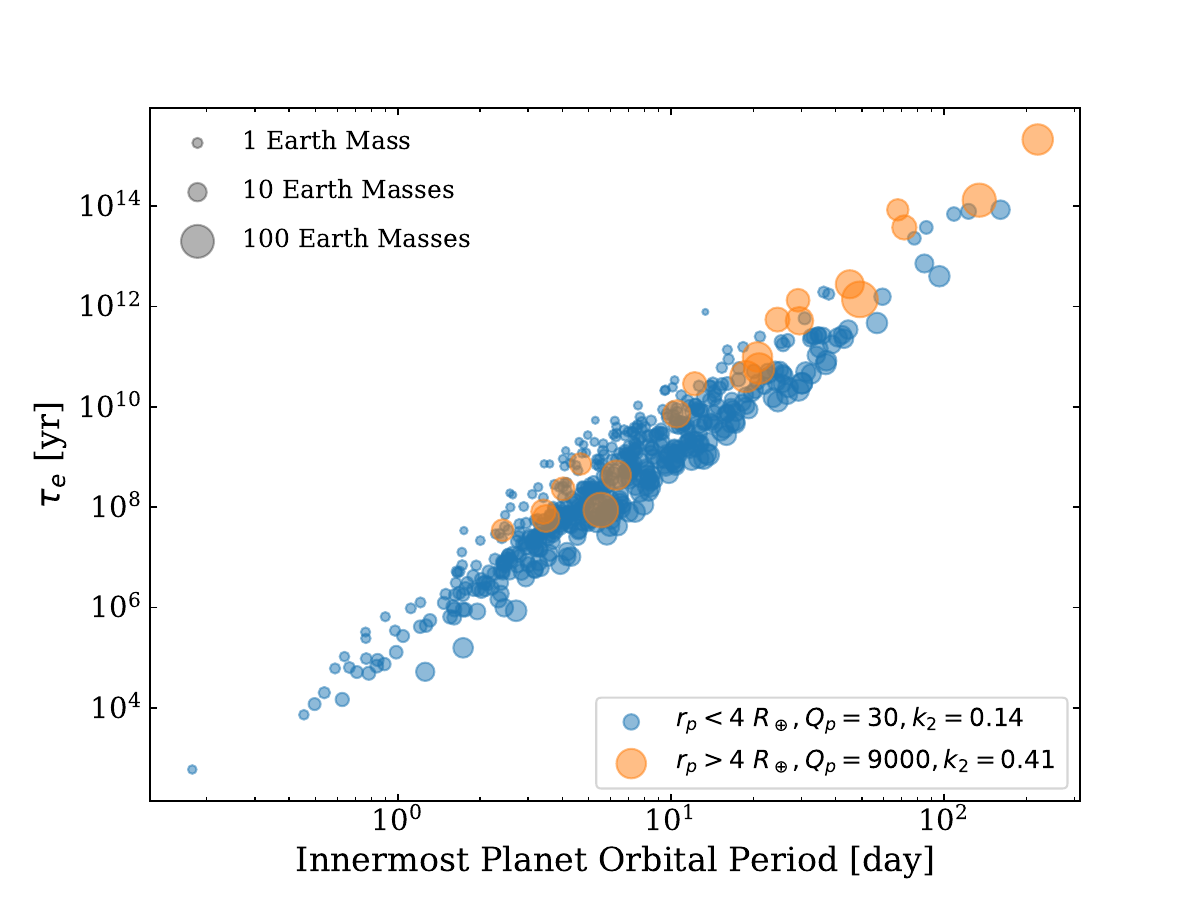}
\caption{Circularization time as a function of orbital period for the
innermost planet in each multiple-planet system in our sample.  We plot
as blue points planets for which we assume the dissipative properties of
Mars $Q_{\text{p}} = 30$ \& $k_{2} = 0.14$ \citep{Duxbury1982,Yoder1995}
and as orange points planets for which we assume the dissipative
properties of Neptune $Q_{\text{p}} = 9000$ \& $k_{2} = 0.41$
\citep{Bursa1992,Zhang2008}.  A point's size encodes the mass of
the planet it represents calculated using the \citet{Lissauer2011}
mass--radius relation.  According to \citet{Lee2013}, tidal evolution
can measurably alter a system's period ratio distribution when the
circularization time of its innermost planet $\tau_{e} \lesssim 20$
Myr.\label{figure03}}
\end{figure*}

We plot in Figure \ref{figure03} circularization time as a function of
orbital period for innermost planets in the Kepler-discovered systems
of multiple planets in our sample.  \citet{Lee2013} showed that the
eccentricities of planet pairs near first-order resonances decay due to
tidal dissipation according to a shallow power law in time.  Those authors
found that more than 50 circularization timescales were necessary to
evolve a system initially in a mean-motion resonance to a period ratio 3\%
away from a rational number.  Because most Kepler-discovered systems of
multiple planets in our sample orbit mature solar-type stars, we assume
that tidal dissipation can only meaningfully alter the near-resonant
period ratios of a system if $50\tau_{e} \lesssim 1$ Gyr or equivalently
if $\tau_{e} < 20$ Myr.

\subsection{Relative Age Inferences}

As argued above, the relative ages of Kepler-discovered systems of
small planets close to/far from period ratios suggestive of mean-motion
resonances have the potential to both (1) decide whether departures from
resonances occur relatively early or relatively late in the evolutionary
histories of exoplanet systems and (2) evaluate the role of tidal
evolution in moving systems out of resonances.  If all multiple-planet
systems are found to be the same age regardless of the period ratios of
their constituent planets, then that observation would suggest that the
features of the period ratio distribution of Kepler-discovered systems
of small planets is set in the first Gyr of planetary system evolution.
If multiple-planet systems with a plausibly resonant planet pair are
younger than multiple-planet systems without a plausibly resonant planet
pair, then planetary systems must evolve out of mean-motion resonances as
they age.  If an age offset is found and restricted to systems with small
innermost planet circularization times, then that would strongly support
tidal evolution as the explanation for the features on the period ratio
distribution of Kepler-discovered systems of small planets.  If an age
offset is found in systems regardless of innermost planet circularization
time, then that would suggest the importance of an additional non-tidal
secular process.

Taking the approach pioneered in \citet{Schlaufman2013} and
\citet{Hamer2019,Hamer2020,Hamer2022}, we use the Galactic velocity
dispersions of these samples to compare their relative ages.  Because the
velocity dispersion of a thin disk stellar population grows with the
average age of that stellar population \citep[e.g.,][]{Binney2000},
populations with significantly different velocity dispersions have
significantly different ages.  The relationship between age and
velocity dispersion in the thin disk of the Milky Way is a function
of the specific region of the Galaxy.  As a result, calibrations like
that of \citet{Binney2000} based on solar neighborhood samples cannot
be directly applied in the Kepler field.  While it is possible to
calibrate the relationship between velocity dispersion and age in the
Kepler field, virtually all stellar age inference techniques ultimately
rely on stellar models.  They are therefore model dependent.  On the
other hand, the \citet{Hamer2019,Hamer2020,Hamer2022} approach enables
precise, model-independent relative age comparisons.  Conclusions based
on model-independent relative ages avoid the systematic uncertainties
associated with other age inference methodologies.  We therefore prefer
these uncalibrated but model-independent relative age inferences for
the analyses that follow.

For a particular sample of stars, we first convert astrometric and
radial velocity data into Galactic $UVW$ space velocities.  For each
star, we next use \texttt{pyia} to generate 100 random sets of
equatorial coordinates, parallaxes, and proper motions in accord with
Gaia astrometric solution covariance matrix for that star.  We then
use \texttt{astropy} to generate 100 random radial velocities in
accord with each star's observed radial velocity and its uncertainty
\citep{astropy2013,astropy2018,PriceWhelan2018}.  We next calculate
100 sets of $UVW$ velocities for each star and then infer 100 Galactic
velocity dispersions for a particular sample according to
\begin{eqnarray}
\sigma = \frac{1}{N}\sum \left[(U_i-\overline{U})^2+(V_i-\overline{V})^2+(W_i-\overline{W})^2\right]^{1/2},
\end{eqnarray}
where $U$, $V$, and $W$ are the components of the Galactocentric velocity
and bars denote median values.  We finally take the median Galactic
velocity dispersion across these 100 realizations as a particular sample's
Galactic velocity dispersion.  The typical $U$, $V$, and $W$ velocity
uncertainties we infer in this way are less than about 0.2 km s$^{-1}$,
much smaller than the typical 1-$\sigma$ ranges of our inferred velocity
dispersion distributions.  Individual stellar $UVW$ velocity uncertainties
are therefore much too small to impact our analyses.

\subsection{Plausibly First-order Resonant Systems}

\begin{figure*}
\centering
\includegraphics[width=\linewidth]{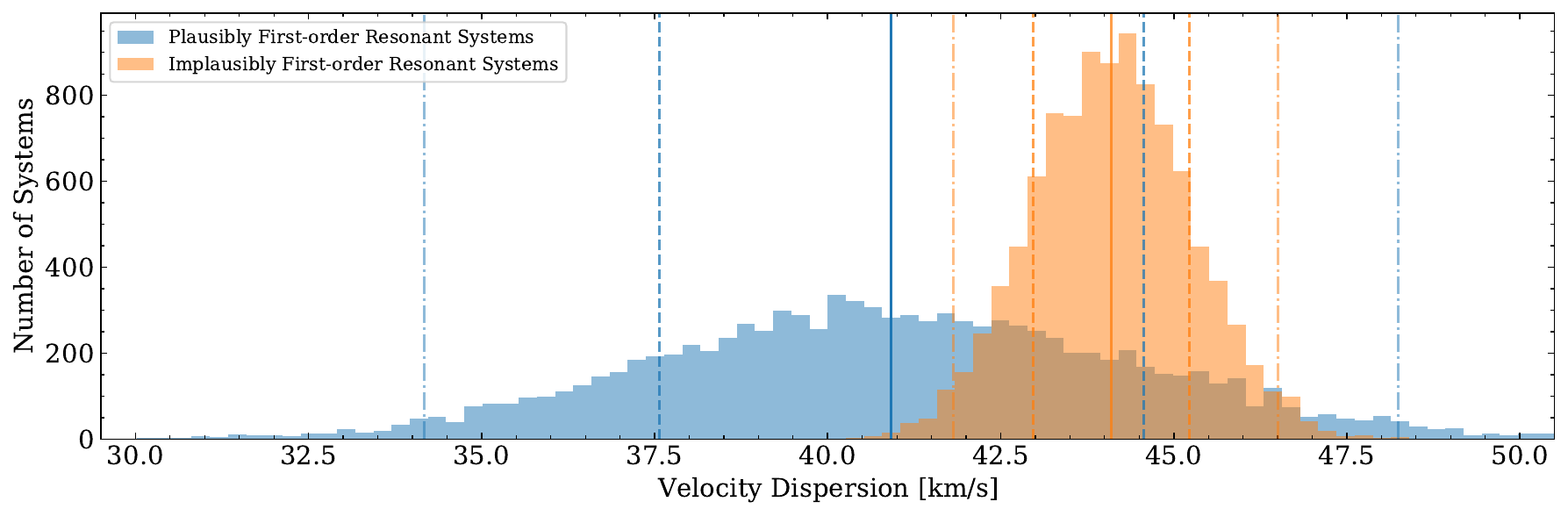}\\
\includegraphics[width=\linewidth]{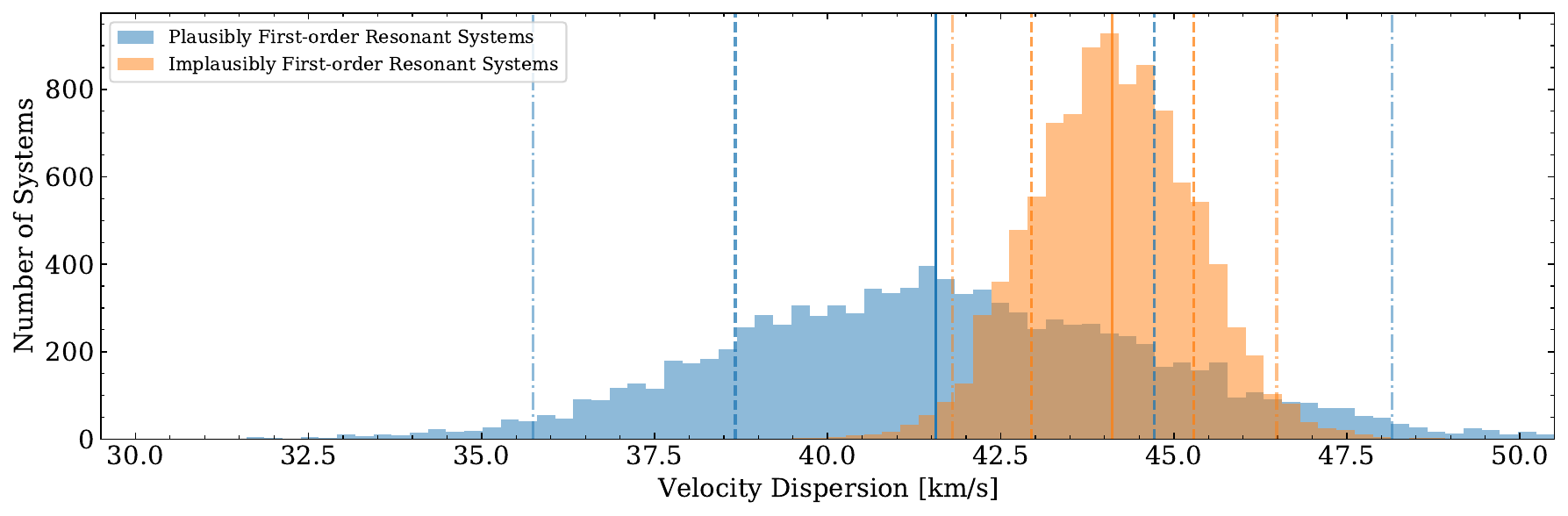}
\caption{Distributions of Galactic velocity dispersion based on bootstrap
resampling for systems with (blue) or without (orange) a plausibly
first-order resonant planet pair.  We indicate median values as solid
vertical lines, 1-$\sigma$ ranges between vertical dashed lines, and
2-$\sigma$ ranges between vertical dot-dashed lines.  Top: plausibly
resonant/implausibly resonant according to the libration width-based
definition.  Bottom: plausibly resonant/implausibly resonant according
to the $\delta_{\text{res}}$-based definition.  The Galactic velocity
dispersion of the sample of Kepler-discovered multiple small-planet
systems with a plausibly first-order resonant planet pair is consistent
with the Galactic velocity dispersion of systems that lack such a pair
regardless of the heuristic used to identify systems with a plausibly
first-order resonant planet pair.\label{figure04}}
\end{figure*}

We first compare the Galactic velocity dispersions and therefore
relative ages of Kepler-discovered systems of small planets with/without
plausibly first-order resonant planet pairs.  This sample of systems
without plausibly first-order resonant planet pairs does include
systems with plausibly second-order resonant planet pairs, though we
have confirmed that our conclusions are unchanged if we also remove
systems with plausibly second-order resonant planet pairs.  We use
5000 bootstrap samples to characterize the underlying Galactic velocity
dispersion distributions of each subsample from the single realization
we observe and plot in Figure \ref{figure04} the resulting velocity
dispersion distributions assuming both the libration width-based and
$\delta_{\text{res}}$-based definitions for plausibly first-order
resonant systems.  We find that systems with plausibly first-order
resonant planet pairs have a Galactic velocity dispersion consistent
with the Galactic velocity dispersion of systems that lack such a pair.
The implication is that the age distribution of systems with plausibly
first-order resonant planet pairs overlaps with the age distribution of
systems that lack such a pair.

\begin{figure*}
\plottwo{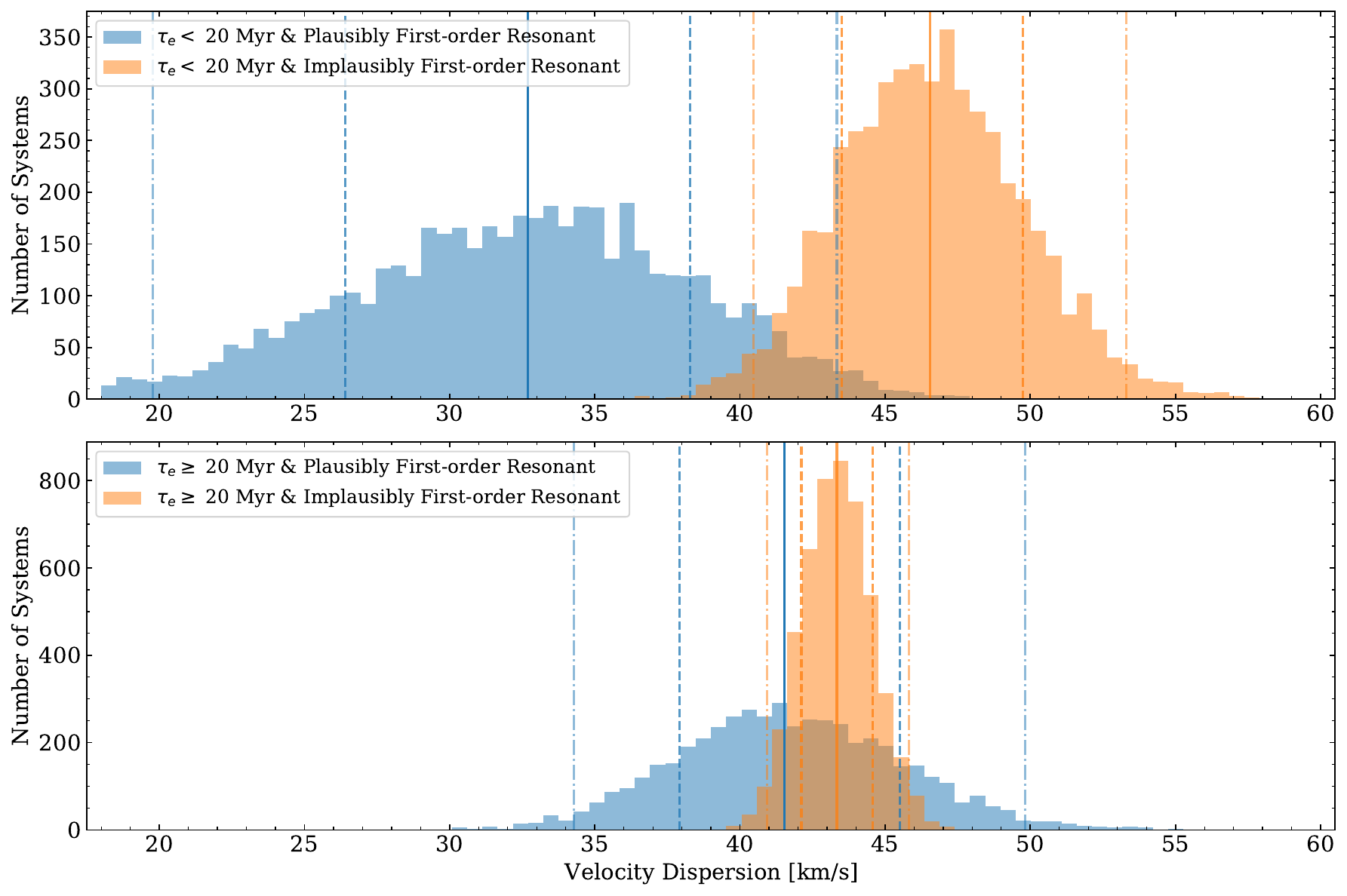}{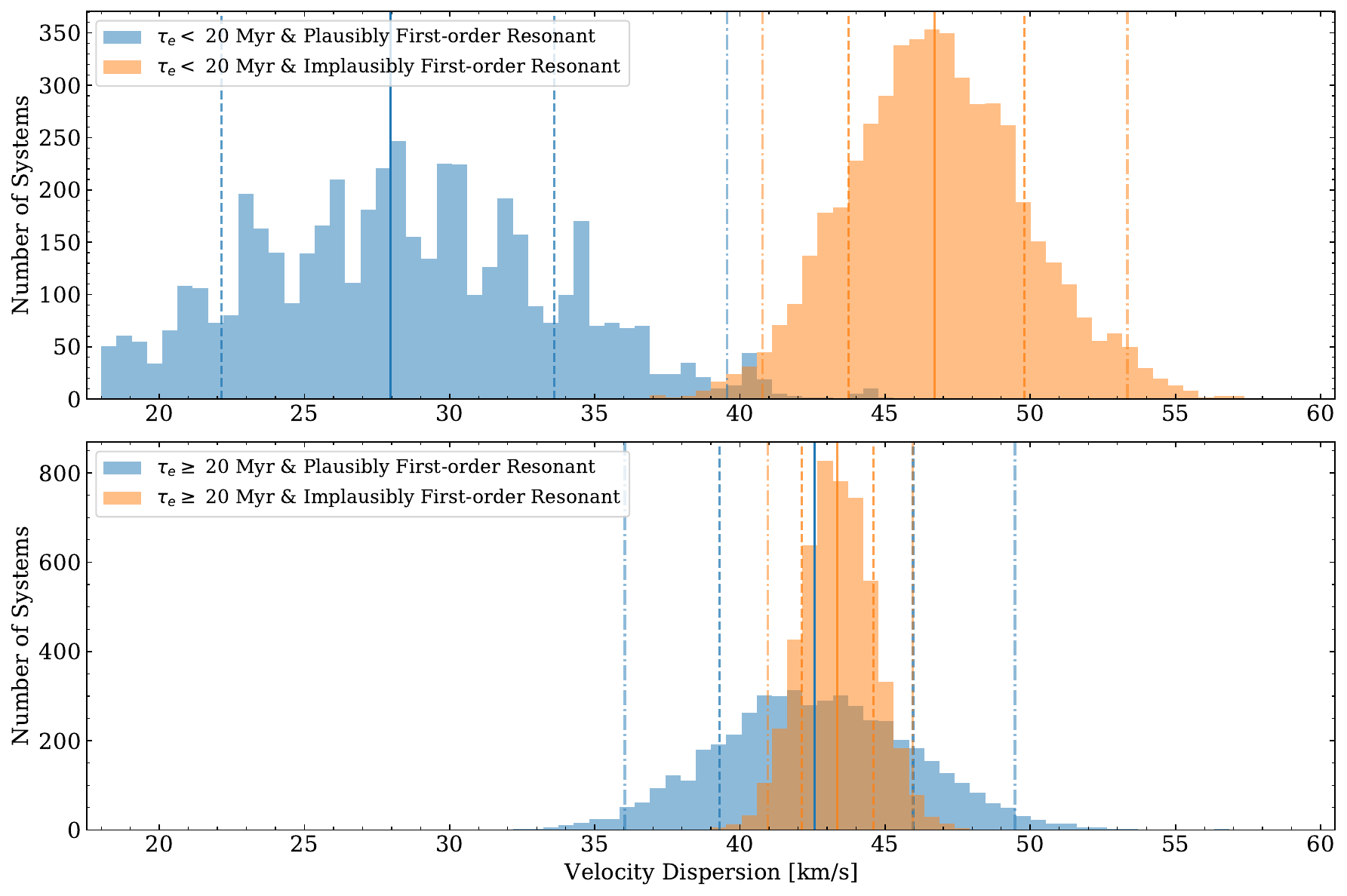}
\caption{Distributions of Galactic velocity dispersion based on bootstrap
resampling for systems with (blue) or without (orange) a plausibly
first-order resonant planet pair.  We indicate median values as solid
vertical lines, 1-$\sigma$ ranges between vertical dashed lines, and
2-$\sigma$ ranges between vertical dot-dashed lines.  Top panels include
systems in which the innermost planet is likely affected by tidal
evolution (i.e., $\tau_{e} < 20$ Myr),  while bottom panels include
systems in which the innermost planet is unlikely to be affected by
tidal dissipation (i.e., $\tau_{e} \geq 20$ Myr).  Left: plausibly
resonant/implausibly resonant according to the libration width-based
definition.  Right: plausibly resonant/implausibly resonant according
to the $\delta_{\text{res}}$-based definition.  The Galactic velocity
dispersion of the sample of Kepler-discovered multiple small-planet
systems likely affected by tides and with a plausibly first-order
resonant planet pair is colder than the Galactic velocity dispersion of
systems that lack such a pair, regardless of the definition of plausibly
first-order resonant systems.  The probabilities that offsets this large
could occur by chance is about 1 in 72 (equivalent to about 2.2 $\sigma$)
for the libration width-based definition and about 1 in 1500 (equivalent
to about 3.1 $\sigma$) for the $\delta_{\text{res}}$-based definition.
The large and statistically significant offsets present in the likely
tidally affected samples support the idea that tidal evolution is
responsible for moving planetary systems formed in first-order resonance
away from those resonances over Gyr timescales.\label{figure05}}
\end{figure*}

We then divide our sample into two subsamples: one in which tidal
dissipation in a system's innermost planet is likely important (i.e.,
$\tau_{e} < 20$ Myr) and one in which such dissipation is unlikely to be
important (i.e., $\tau_{e} \geq 20$ Myr).  We apply the same bootstrap
resampling strategy described in the preceding paragraph and plot in
Figure \ref{figure05} the resulting velocity dispersion distributions.
For systems with an innermost planet likely affected by tides, we find
that systems with a plausibly first-order resonant planet pair have a
colder Galactic velocity dispersion than systems that lack a plausibly
resonant planet pair.  This is so for both the libration width-based and
$\delta_{\text{res}}$-based definitions for plausibly first-order resonant
systems.  For the libration width-based definition, the probability that
an offset this large could be produced by chance is about 1 in 72 or
equivalently about 2.2 $\sigma$.  For the $\delta_{\text{res}}$-based
definition, the probability that an offset this large could be produced
by chance is about 1 in 1500 or equivalently about 3.1 $\sigma$.
These observations suggest that that tidal dissipation plays an important
role in moving away from first-order resonances multiple-planet systems
left in first-order resonances after the dissipation of their parent
protoplanetary disks.

\begin{figure*}
\plottwo{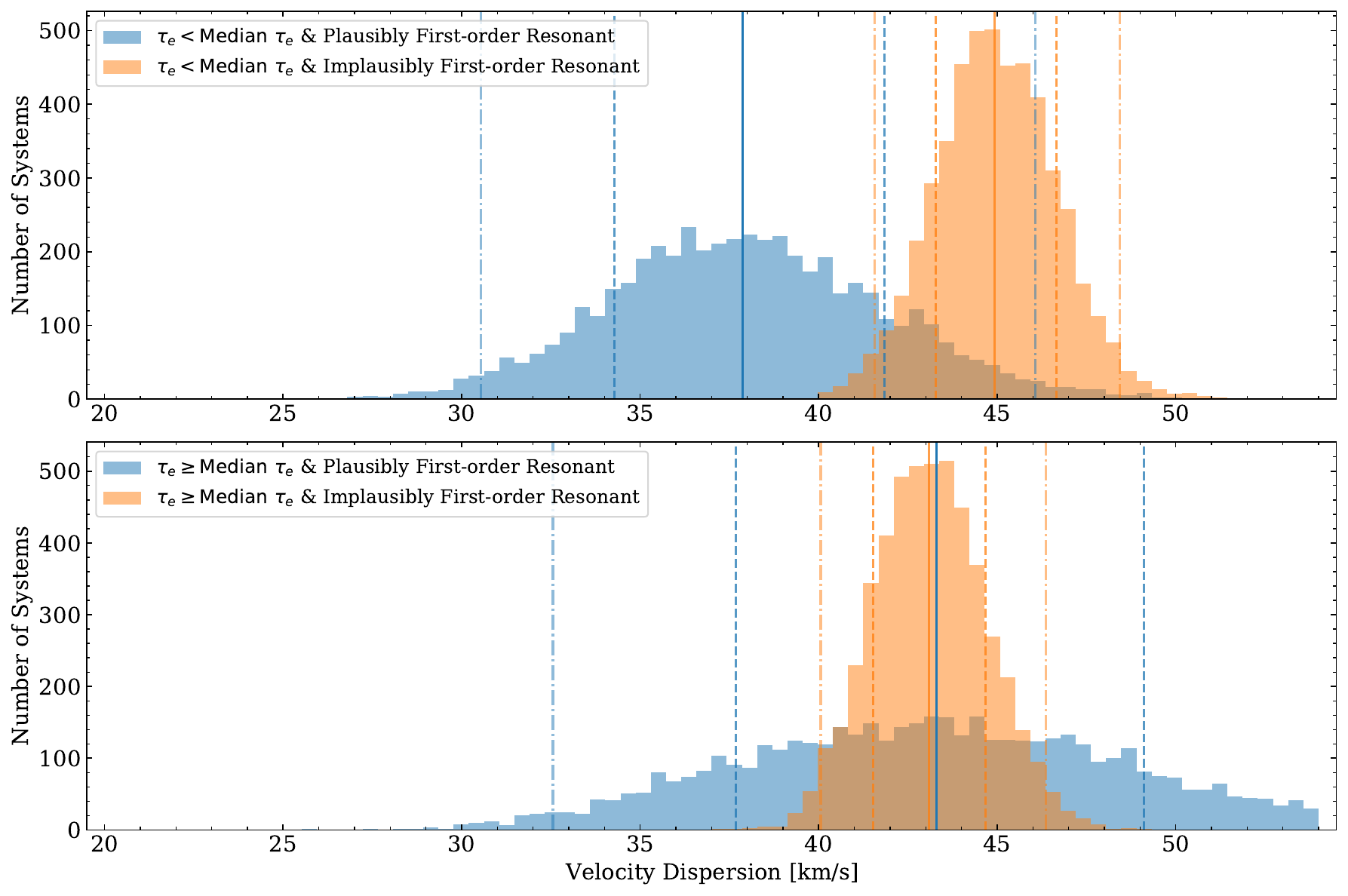}{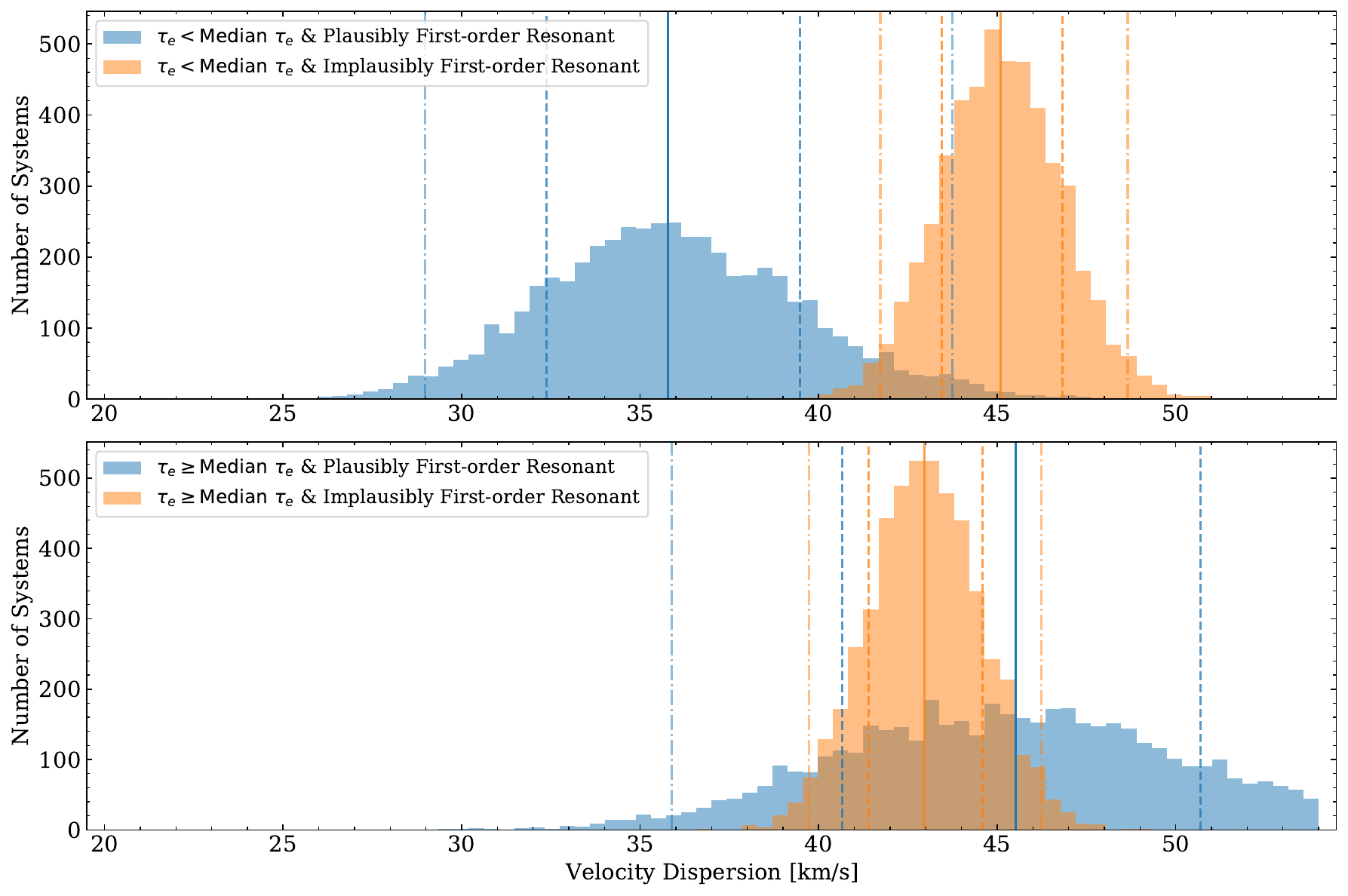}
\caption{Distributions of Galactic velocity dispersion based on
bootstrap resampling for systems with (blue) or without (orange)
a plausibly first-order resonant planet pair.  We indicate median
values as solid vertical lines, 1-$\sigma$ ranges between vertical
dashed lines, and 2-$\sigma$ ranges between vertical dot-dashed lines.
Top panels include systems in which the innermost planet is likely
affected by tidal evolution (i.e., $\tau_{e} < \overline{\tau}_{e}
= 400$ Myr),  while bottom panels include systems in which the
innermost planet is unlikely to be affected by tidal dissipation
(i.e., $\tau_{e} \geq \overline{\tau}_{e} = 400$ Myr).  Left: plausibly
resonant/implausibly resonant according to the libration width-based
definition.  Right: plausibly resonant/implausibly resonant according
to the $\delta_{\text{res}}$-based definition.  Regardless of the
definition used to identify plausibly first-order resonant systems,
systems with innermost planet circularization times less than the median
value $\overline{\tau}_{e} = 400$ Myr have a colder Galactic velocity
dispersion and are therefore younger than implausibly resonant systems.
The probabilities that offsets in the tidally affected samples this
large could occur by chance are about 1 in 18 (equivalent to about 1.6
$\sigma$) for the libration-width definition and 1 in 93 (equivalent
to about 2.3 $\sigma$) for the $\delta_{\text{res}}$-based definition.
Our conclusion that plausibly first-order resonant systems with an
innermost planet likely to be affected by tidal dissipation are young
does not sensitively depend on the criteria used to identify systems
likely affected by tides.\label{figure06}}
\end{figure*}

It has been proposed in certain models accounting
for secular interactions in compact multiple-planet
systems \citep[e.g.,][]{Hansen2015} or obliquity tides
\citep[e.g.,][]{Millholland2019a,Millholland2019b,Millholland2019c}
that the efficiency of tidal dissipation inside an innermost planet may
exceed our assumed value.  To investigate this possibility, we divided
our sample based on its median innermost planet circularization time
$\overline{\tau}_{e} = 400$ Myr.  Dividing the sample at its median
value allows for a qualitative examination of the role of tides
independent of a particular model while also equalizing the sizes of
the tidal/non-tidal samples and maximizing the statistical power of
their comparison.  We report the outcome of this exercise in Figure
\ref{figure06}.  Regardless of the definition used to identify plausibly
resonant planet pairs, systems with circularization times $\tau_{e} <
\overline{\tau}_{e}$ and a plausibly first-order resonant planet pair
have colder Galactic velocity dispersion than systems that lack such
a pair.  The probabilities that offsets in the tidally affected samples
this large could occur by chance are about 1 in 18 (equivalent to about
1.6 $\sigma$) for the libration-width definition and 1 in 93 (equivalent
to about 2.3 $\sigma$) for the $\delta_{\text{res}}$-based definition.
On the other hand, systems with circularization times $\tau_{e} >
\overline{\tau}_{e}$ do not display an offset in Galactic velocity
dispersion between plausibly first-order resonant/implausibly first-order
resonant systems.  The similarity between our results using two different
definitions for the importance of tidal evolution shows that our results
are not sensitively dependent on the choice of tidal parameters.

\subsection{Plausibly Second-order Resonant Systems}

\begin{figure*}
\plotone{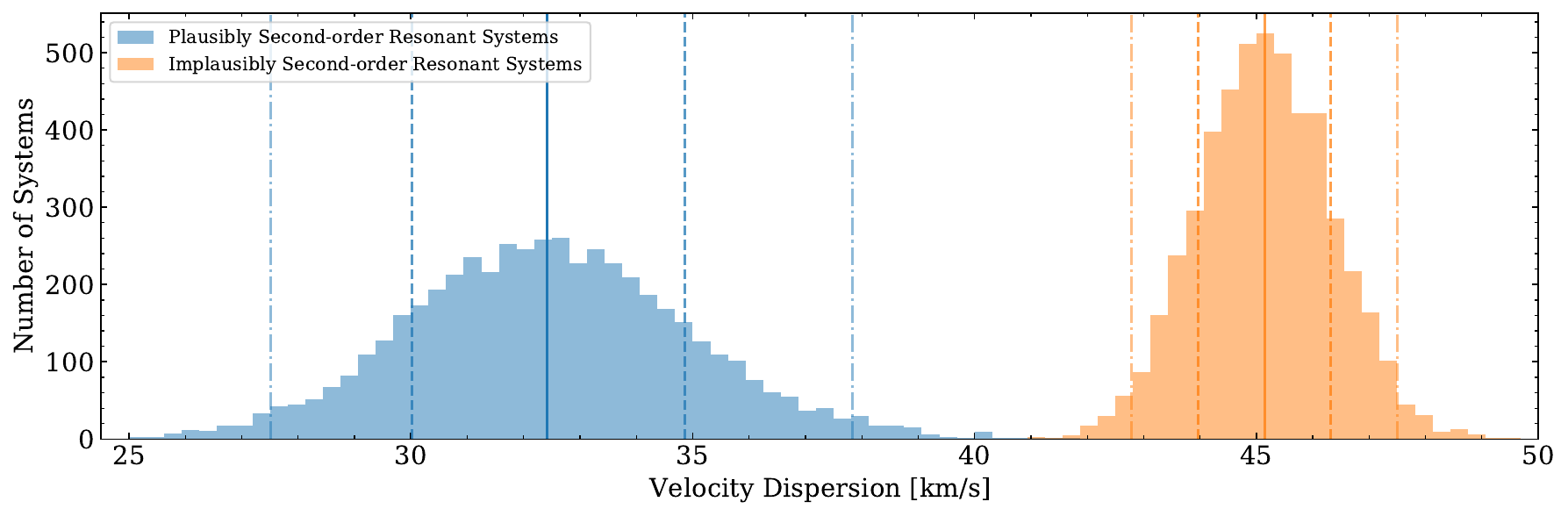}
\caption{Distribution of Galactic velocity dispersion based on
bootstrap resampling for systems with (blue) or without (orange)
a plausibly second-order resonant planet pair according to the
$\delta_{\text{res}}$-based definition.  We indicate median values as
solid vertical lines, 1-$\sigma$ ranges between vertical dashed lines,
and 2-$\sigma$ ranges between vertical dot-dashed lines.  The Galactic
velocity dispersion of the sample of Kepler-discovered multiple
small-planet systems with a plausibly second-order resonant planet pair is
significantly colder than the Galactic velocity dispersion of systems that
lack such a pair.  The probability that an offset this large could occur
by chance is less than 1 in 100000, or equivalently about 4.6 $\sigma$.
This observation implies that systems initially in second-order resonance
are driven out of resonance over Gyr timescales.\label{figure07}}
\end{figure*}

We next compare the Galactic velocity dispersions and therefore relative
ages of Kepler-discovered systems of small planets with/without plausibly
second-order resonant planet pairs using the same procedure described
in the preceding subsection.  This sample of systems without plausibly
second-order resonant planet pairs does include systems with plausibly
first-order resonant planet pairs, though we have confirmed that our
conclusions are unchanged if we also remove systems with plausibly
first-order resonant planet pairs.  The libration-width based definition
is only applicable for plausibly first-order resonant systems, so
we only use the $\delta_{\text{res}}$-based definition for plausibly
second-order resonant systems in this subsection.  We plot in Figure
\ref{figure07} the resulting velocity dispersion distributions under
the $\delta_{\text{res}}$-based definition for plausibly second-order
resonant systems.  We find that systems with plausibly second-order
resonant planet pairs have a significantly colder Galactic velocity
dispersion than systems that lack such a pair.  The implication is that
systems with plausibly second-order resonant planet pairs are younger
than systems that lack such a pair.  The probability that an offset this
large could occur by chance is less than 1 in 100000, or equivalently
about 4.6 $\sigma$.

\begin{figure*}
\plottwo{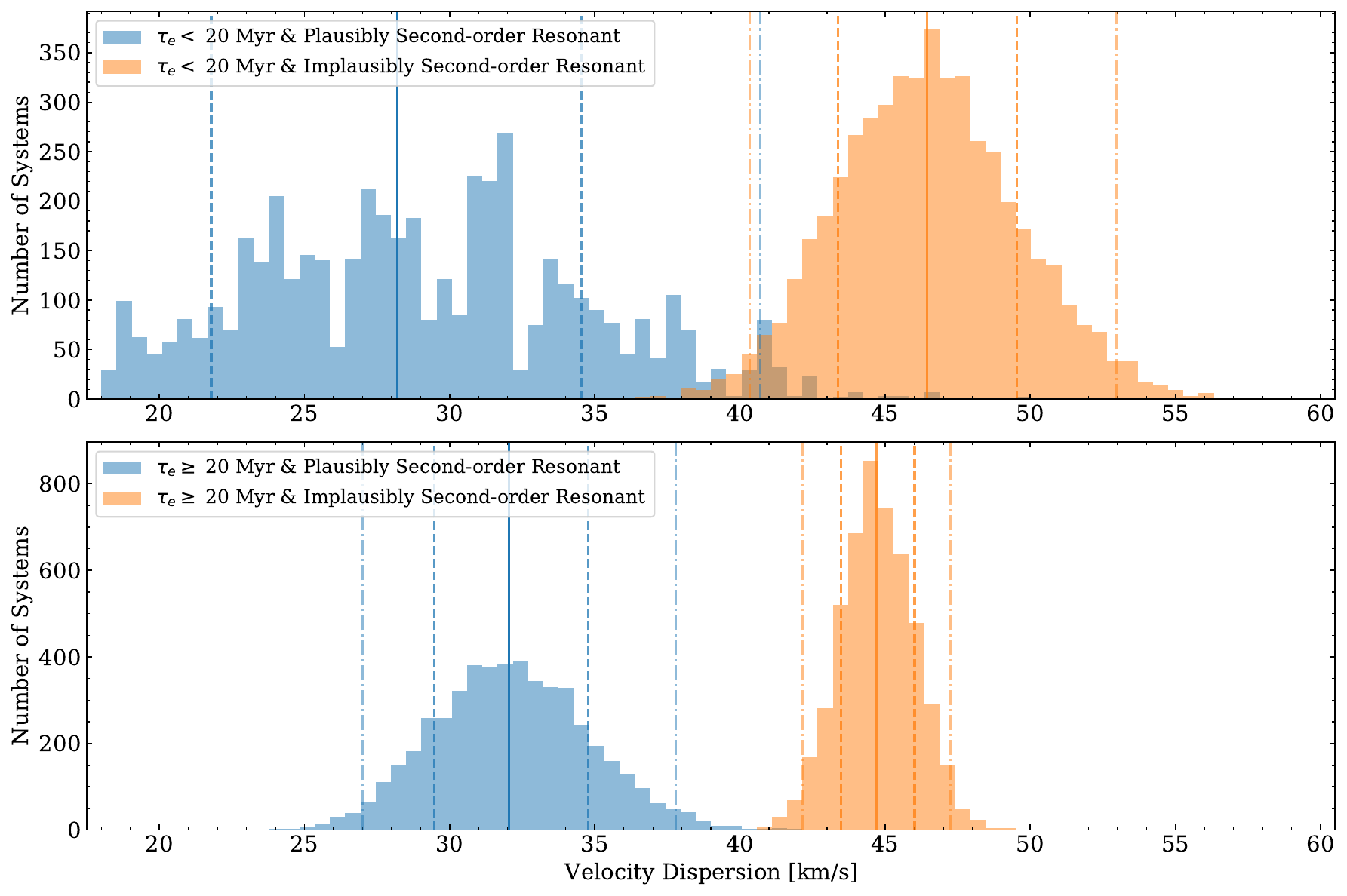}{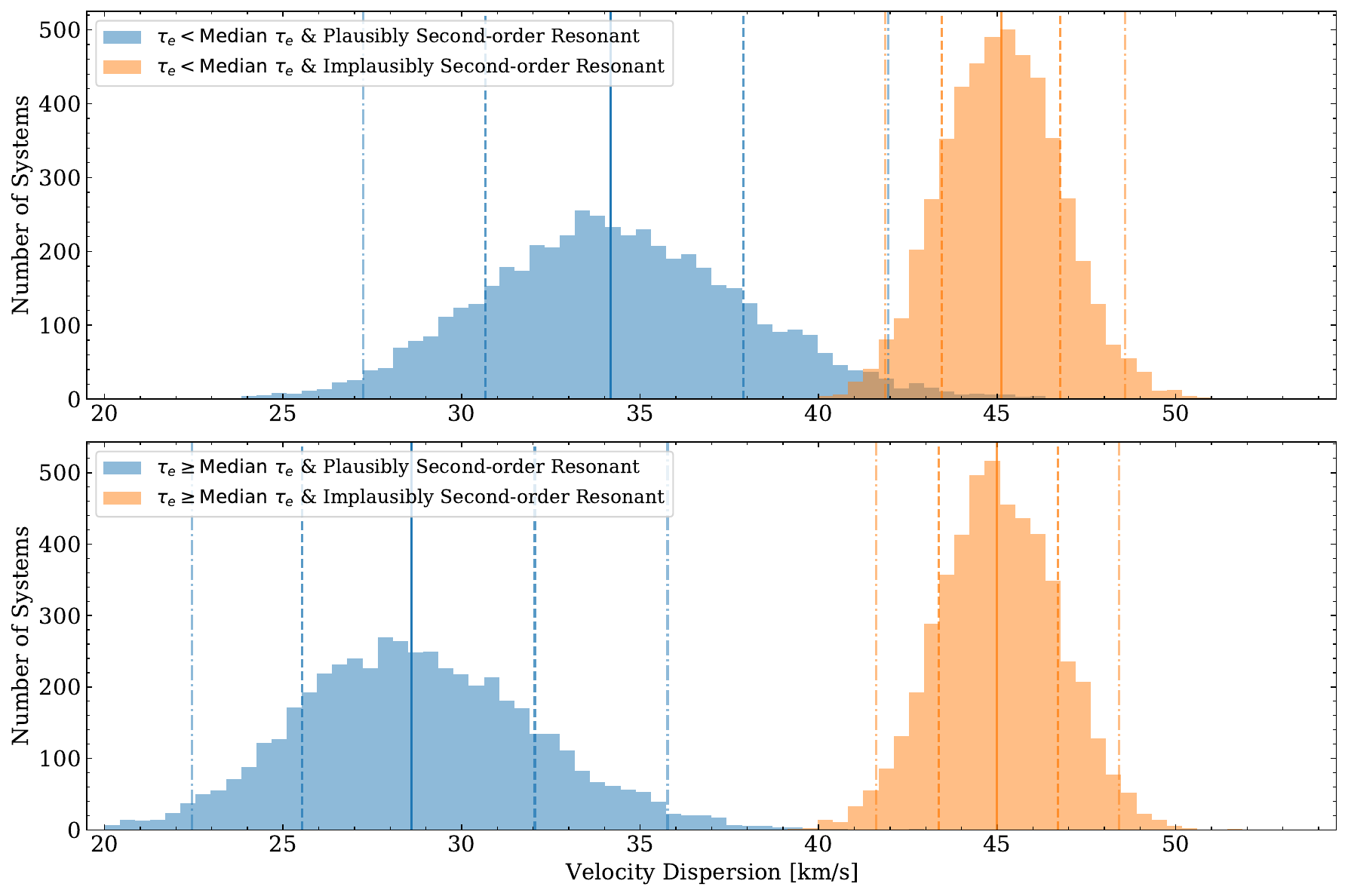}
\caption{Distributions of Galactic velocity dispersion based on
bootstrap resampling for systems with (blue) or without (orange)
a plausibly second-order resonant planet pair according to the
$\delta_{\text{res}}$-based definition.  We indicate median values as
solid vertical lines, 1-$\sigma$ ranges between vertical dashed lines,
and 2-$\sigma$ ranges between vertical dot-dashed lines.  Top panels
include systems in which the innermost planet is likely affected by
tidal evolution,  while bottom panels include systems in which the
innermost planet is unlikely to be affected by tidal dissipation.
Left: Tidal and non-tidal samples separated at $\tau_{e} = 20$ Myr.
The probabilities that offsets this large could occur by chance is about 1
in 210 (equivalent to about 2.6 $\sigma$) for the systems likely affected
by tides and about 1 in 76000 (equivalent to about 4.2 $\sigma$) for
systems unlikely to be unaffected by tides.  Right: Tidal and non-tidal
samples separated by the median circularization time $\tau_{e} = 400$ Myr.
The probabilities that offsets this large could occur by chance is about 1
in 280 (equivalent to about 2.7 $\sigma$) for the systems likely affected
by tides and about 1 in 76000 (equivalent to about 4.2 $\sigma$) for
systems unlikely to be unaffected by tides.  Regardless of the importance
of tidal dissipation inside innermost planets, the Galactic velocity
dispersion of the sample of Kepler-discovered multiple small-planet
systems with a plausibly second-order resonant planet pair is colder
than the Galactic velocity dispersion of systems that lack such a pair.
The implication is that tidal dissipation is not the process responsible
for moving systems initially in second-order mean-motion resonance out
of resonance.\label{figure08}}
\end{figure*}

To investigate the possible importance of tidal evolution in moving
systems away from second-order resonance, as before we split our sample
of systems into two subsamples: one in which tidal dissipation inside
the innermost planet is likely important (i.e., $\tau_{e} < 20$ Myr or
$\tau_{e} < \overline{\tau}_{e} = 400$ Myr) and one in which tides are
unlikely to be important (i.e., $\tau_{e} \geq 20$ Myr or $\tau_{e}
\geq \overline{\tau}_{e} = 400$ Myr).  We apply the same bootstrap
resampling strategy described in the preceding subsection and plot in
Figure \ref{figure08} the resulting velocity dispersion distributions.
We find that systems with a plausibly second-order resonant planet pair
have a significantly colder Galactic velocity dispersion than systems that
lack such a pair regardless of the possible impact of tidal dissipation.
When splitting our sample at $\tau_{e} = 20$ Myr, the probabilities of
offsets as large as we observe by change are about 1 in 210 (equivalent
to about 2.6 $\sigma$) and about 1 in 76000 (equivalent to about 4.2
$\sigma$) for the systems affected/unaffected by tides.  When splitting
our sample at $\overline{\tau}_{e} = 400$ Myr, the probabilities of
offsets as large as we observe by change are about 1 in 280 (equivalent
to about 2.7 $\sigma$) and about 1 in 76000 (equivalent to about 4.2
$\sigma$) for the systems affected/unaffected by tides.

For systems with a plausibly second-order resonant planet pair, the lack
of a dependence of velocity dispersion offset on the exact criteria used
to identify systems affected by tidal evolution suggests that tidal
dissipation is not the best candidate to explain our observation that
systems with a plausibly second-order resonant planet pair are younger
than systems that lack such a pair.  As we will argue in the next section,
some other non-tidal secular processes is likely responsible for drawing
systems out of second-order resonances.

\section{Discussion}

The Galactic velocity dispersion of the sample of Kepler-discovered
multiple-planet systems with at least one plausibly second-order resonant
planet pair is colder than the Galactic velocity dispersion of the
sample of Kepler-discovered multiple-planet systems without such a pair.
The implication is that systems with a plausibly second-order resonant
pair are systematically young.  The same is true for Kepler-discovered
multiple-planet systems with at least one plausibly first-order resonant
planet pair, but only when the innermost planet in the system is close
enough to its host that tidal dissipation inside the planet likely
affects the secular evolution of the system.

\subsection{Evaluation of Potentially Observable Consequences of the
Tidal Evolution Implied by Our Observations}

Because tides appear to be responsible for the Galactic velocity
dispersion and therefore age offsets we observe between first-order
resonant and non-resonant systems, we now assess whether or not the
tidal heating of innermost planets in first-order resonant systems with
forced eccentricities would be observable.  Following \citet{Jackson2008}
the tidal heating power experienced by a planet can be written
\begin{eqnarray}
H = \frac{63}{4} \frac{(G M_{\ast})^{3/2} M_{\ast}
R_{p}^{5}}{Q'_p} a^{-15/2} e^{2},
\end{eqnarray}
where $G$ is the gravitational constant, $M_{\ast}$ its host star's mass,
$R_{p}$ its radius, $Q'_{p}$ its modified tidal quality factor, $a$
its semimajor axis, and $e$ its orbital eccentricity.  For the sample
of Kepler-discovered multiple-planet systems we identified as plausibly
first-order resonant, non-zero innermost planet orbital eccentricities
and Equation (11) can be used to determine the tidal heating experienced
by innermost planets.  We use the same $Q'_p$ values we assumed in out
circularization timescale calculations and assume $e = 0.02$ corresponding
to the median libration width-minimizing eccentricity in our sample.
We find a typical tidal heating power $H \sim 10^{15}$ W and a maximum
$H \gtrsim 10^{17}$ W.

While important for the internal structures of rocky planets, the
observable effects of even $H \gtrsim 10^{17}$ W of tidal heating
are unlikely to be observable.  \citet{Valencia2007} showed that
$10^{17}$ W/$M_\oplus$ of tidal heating can cause the partial melting
of a planet's lower lower mantle, and we identify four planets
in our sample experiencing tidal heating exceeding that threshold:
Kepler-1371 c/KOI-2859.02 ($H \approx 2.6 \times 10^{17}$ W), Kepler-1669
c/KOI-1475.01 ($H \approx 3.2 \times 10^{17}$ W), Kepler-9 d/KOI-377.03
($H \approx 3.7 \times 10^{17}$ W), and Kepler-342 e/KOI-1955.03 ($H
\approx 5.2 \times 10^{17}$ W).  \citet{Valencia2007} also showed that
$H \approx 6.8 \times 10^{17}$ W of tidal heating in GJ 876 d would only
increase its radius by about 100 km.  As a result, in the four systems
listed above the observational consequences of the tidal heating are
much smaller than the observational consequences of uncertainties in
mass--radius relations due to planet composition.  They are therefore
unobservable as larger-than-expected planet radii.

These tidal heating powers are also unlikely to be observable in the form
of higher-than-expected planet equilibrium temperatures.  For a possibly
measurable effect, the tidal heating experienced by a planet would need to
be comparable to its instellation flux multiplied by its cross-sectional
area and one minus its Bond albedo.  Only planets with small orbital
separations experience significant tidal heating, and these plants also
experience intense instellations that are much more important for setting
their equilibrium temperatures than tidal heating.  On the other hand,
planets in high planetary obliquity states affected by obliquity tides
may experience tidal heating with sufficient intensities to increase
their equilibrium temperatures \citep[e.g.,][]{Millholland2019a}.
An observation that revealed an enhanced planetary equilibrium temperature
would be hard to confidently attribute to obliquity tides though, as
apparently increased equilibrium temperatures could be caused by an
overestimated planetary Bond albedo or by the impact of an atmospheric
greenhouse effect.  A possibly more easily observable consequence of
intense tidal heating from obliquity tides would by inflated radii of
planets with significant hydrogen/helium atmospheres wide of mean-motion
resonances \citep[e.g.,][]{Millholland2019a,Millholland2019c}.

\subsection{Consistency Between our Results and Published Explanations
for the Period Ratio Distribution Observed by Kepler}

\begin{figure*}
\plotone{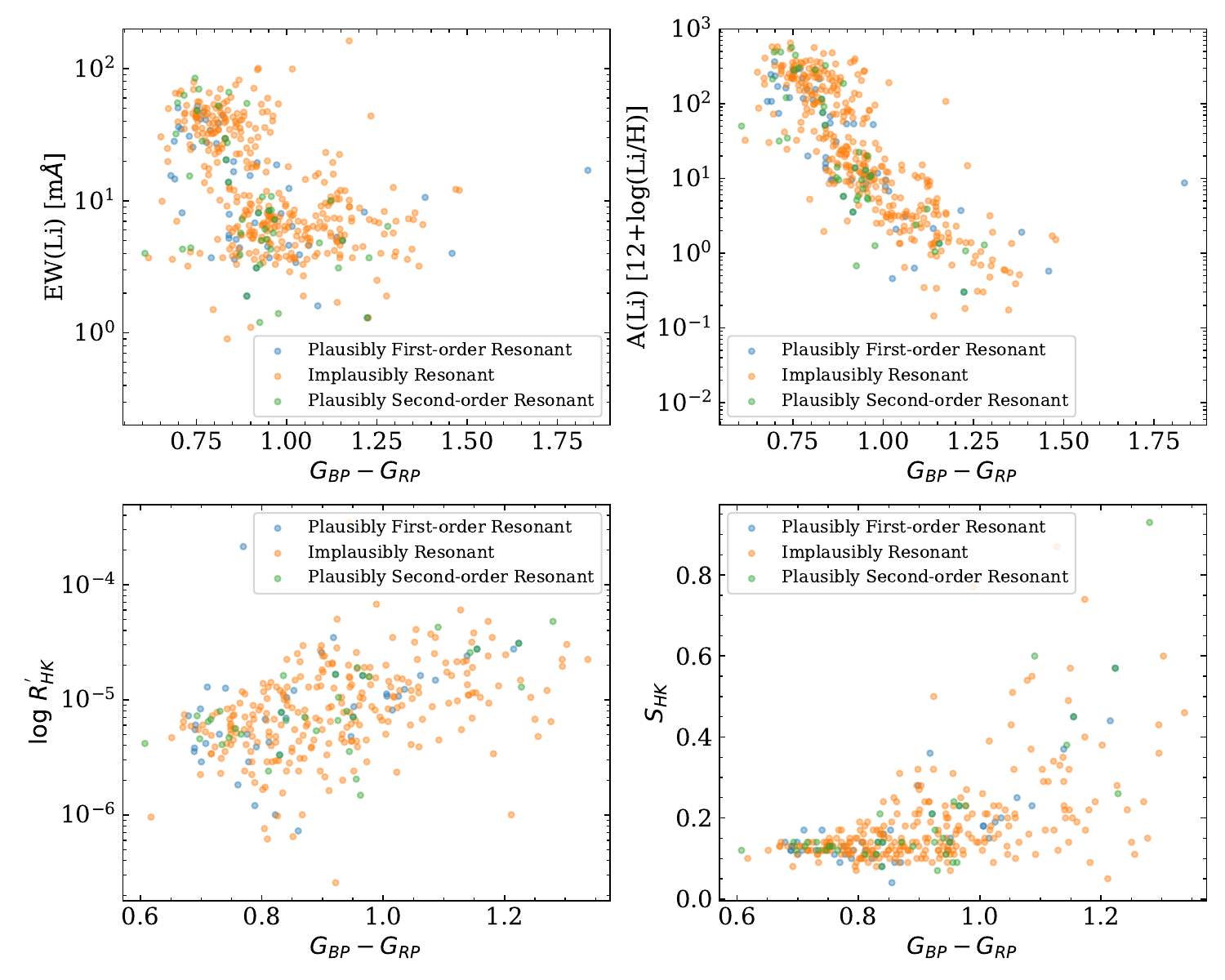}
\caption{Indirect age indicators for our sample of Kepler multiple-planet
system host stars.  We plot as blue points systems with at least one
plausibly resonant planet pair and as orange points systems that lack a
plausibly resonant planet pair.  Top left: from \citet{Berger2018} the
equivalent width in m\AA~of the 670.8 nm lithium feature as a function
of $G_\text{BP}-G_\text{RP}$.  Top right: from \citet{Berger2018}
lithium abundance as a function of $G_\text{BP}-G_\text{RP}$.
At constant $G_\text{BP}-G_\text{RP}$, large equivalent widths
or lithium abundances should be observed in the spectra of stars
younger than about 650 Myr.  We see no significant equivalent width
or abundance offsets between plausibly resonant/implausibly resonant
systems, suggesting that these plausibly resonant systems are not
young in an absolute sense.  Bottom left: from \citet{Brewer2018}
$\log{R'_{\text{HK}}}$ as a function of $G_\text{BP}-G_\text{RP}$.
Bottom right: from \citet{Brewer2018} $S_{\text{HK}}$ as a function of
$G_\text{BP}-G_\text{RP}$.  $\log{R'_{\text{HK}}}$ and $S_{\text{HK}}$
are measures of the strength of emission in the core of the calcium H \&
K lines.  At constant $G_\text{BP}-G_\text{RP}$, stronger emission should
be observed in the spectra of younger stars.  We see no significant
$\log{R'_{\text{HK}}}$ or $S_{\text{HK}}$ offsets between plausibly
resonant/implausibly resonant systems, suggesting that these plausibly
resonant systems are not young in an absolute sense.\label{figure09}}
\end{figure*}

Our observation of a velocity dispersion and therefore age offset between
our samples of plausibly second-order resonant and implausibly resonant
systems unlikely to be affected by tidal evolution suggests that some
other process(es) beyond tidal dissipation may be necessary to explain our
observation.  While our observation has no implications for the absolute
ages of plausibly second-order resonant/implausibly resonant systems,
the Galactic velocity dispersion offset we observe implies a relative
age difference between plausibly second-order resonant/implausibly
resonant systems of at least 100 Myr.  To investigate the possibility
that plausibly resonant systems are a young population with $\text{age}
\lesssim 1$ Gyr and implausibly resonant systems are an old population
with $\text{age} \gtrsim 1$ Gyr, we examine the distribution of
lithium equivalent width and abundance as well as \ion{Ca}{2} H and
K $\log{R'_{\text{HK}}}$ and $S_\text{HK}$ activity indices in both
populations using data from \citet{Berger2018} and \citet{Brewer2018}.
We plot these distributions in Figure \ref{figure09}.  Neither population
displays obvious signs of youth and plausibly resonant systems do not have
obviously larger observed lithium equivalent widths, inferred lithium
abundances, or $\log{R'_{\text{HK}}}$/$S_\text{HK}$ activity indices.
The implication is that neither population has a typical $\text{age}
\lesssim 1$ Gyr.

Because the host stars of both plausibly second-order resonant and
implausibly resonant systems have lithium properties and activity indices
suggesting ages in excess of 100 Myr, we assert that it is unlikely that
the age difference between the two populations we identify is imparted
by a process driving second-order resonant planet pairs out of resonance
within the first 100 Myr of a system's evolution.  Indeed, if initially
resonant systems had already been driven away from resonance by the
time of their parent protoplanetary disks' dissipations, then the age
difference between plausibly resonant/implausibly resonant systems would
be small and significant Galactic velocity dispersion offsets between
populations would not exist.  We therefore argue that in situ formation
of planets \citep[e.g.,][]{Petrovich2013}, planetary interactions
with protoplanetary disk density waves raised by additional planets
\citep[e.g.,][]{PodlewskaGaca2012,Baruteau2013,Cui2021}, planet--planet
interactions in the presence of protoplanetary disk-mediated
eccentricity damping \citep[e.g.,][]{Charalambous2022,Laune2022}, and
system disruptions due to progressive protoplanetary disk dissipation
\citep[e.g.,][]{Liu2022} are all unlikely to explain the multiple-planet
system period ratio distribution observed by Kepler.

Because most Kepler-discovered multiple-planet systems orbit solar-type
stars that spin down on the order of 100 Myr, we also assert that evolving
host star quadrupolar potentials are unlikely to be responsible for
disrupting resonant systems and generating the age offset we inferred.
Stellar oblateness due to rotation can cause small separation planetary
systems to depart from idealized Keplerian orbits.  In these cases,
the quadrupole term of the stellar potential that scales like $J_{2}
R_{\ast}^2$ is usually dominant \citep[e.g.,][]{Schultz2021}.  The stellar
quadrupole moment $J_{2}$ can be approximated as
\begin{eqnarray}
J_{2} & \approx & \frac{1}{3} k_2 \left(\frac{\Omega_{\ast}^2 R_{\ast}^{3}}{GM_{\ast}}\right),
\end{eqnarray}
where $k_2$ is related to the apsidal motion constant $k_2/2$ and
$\Omega_{\ast}$, $M_{\ast}$, and $R_{\ast}$ are the stellar angular
velocity, mass, and radius \citep[e.g.,][]{Schultz2021}.  We use the
stellar models of \citet{Ekstrom2012} that account for the rotational
evolution of stars to evaluate the time evolution of $J_{2} R_{\ast}^2$.
We take $k_2$ from \citet{Batygin2013a} assuming apsidal motion constants
$k_2/2$ \citep[e.g.,][]{Csizmadia2019} as they do for fully convective
($k_{2}/2 = 0.14$) and fully radiative stars ($k_{2}/2=0.01$).  Because a
solar mass and composition main sequence star has a radiative core and
a convective envelope, we assume $k_2 = 0.14$ for the limiting case
where evolution of the quadrupole term of the stellar potential will be
most important.  We find that $J_{2} R_{\ast}^2$ evolves by an order of
magnitude during the first 100 Myr, by a factor of several during next
few hundred Myr, and by very little after 1 Gyr for the duration of the
main sequence.  We therefore argue that evolving stellar quadrupolar
potentials are unlikely to explain the multiple-planet system period
ratio distribution observed by Kepler.

It has been shown that initially resonant multiple-planet systems
can become unstable shortly after the dissipations of their parent
protoplanetary disks with instability times inversely proportional
to the number of planets in a system and their characteristic mass
\citep[e.g.,][]{Matsumoto2012,Izidoro2017,Pichierri2018,Pichierri2020,Izidoro2021,Goldberg2022a,Goldberg2022b,Rice2023}.
Instabilities can also arise in initially stable systems as a result of
either stellar or planetary mass loss.  Tightly packed resonant systems
close to the threshold for dynamical instability can become unstable
if host stars loses just 1\% of their masses or if systems' total
planetary masses change by more than 10\% \citep[e.g.,][]{Matsumoto2020}.
Larger mass losses can produce instabilities even in systems initially
far from the threshold for dynamical instability.  \citet{Wood2002}
derived a relationship between observed X-ray fluxes and stellar mass
loss rates and uses that relationship to predict the mass loss rates of
solar-type stars as function of stellar age.  They made the provocative
suggestion that the Sun may have lost up to 3\% of its mass by 1 Gyr.
Even if this extreme estimate is accurate, the vast majority of that
mass loss takes place in the first 100 Myr of a system's evolution.
That timescale is too short to explain our observations.  Late time
dynamical instabilities in multiple-planet systems driven by stellar mass
loss are therefore unlikely to be responsible for disrupting resonant
systems and generating the age offset we inferred.

Late time dynamical instabilities in multiple-planet systems driven by
planetary mass loss may play some role in the disruption of initially
resonant systems and the production of the age offsets we inferred though.
To evaluate this scenario for the typical system in our sample, we use
the models of \citet{Lopez2013}.  The typical planet in our sample has
$M_{\text{p}} \approx 5~M_{\oplus}$ and experiences an instellation
$F_{\text{p}} \sim 100~F_{\oplus}$.  Assuming the \citet{Lopez2013}
preferred mass loss efficiency factor equal to 0.1, the typical planet
in our sample that initially possessed a massive hydrogen/helium (H/He)
atmosphere could have lost more than 80\% of its mass in H/He in just
50 Myr.  A planet that originally had just 10\% of its mass in H/He would
lose its entire atmosphere in 100 Myr.  Both timescales are too short
explain our result.  On the other hand, a planet with 10\% of its original
mass in H/He experiencing $F_{\text{p}} \approx 50~F_{\oplus}$ would lose
50\% of this mass in 100 Myr and the rest in 1 Gyr.  At face value then,
dynamical instabilities due to the loss of primordial H/He atmospheres
in initially resonant systems experiencing intermediate instellations
could play some role in the explanation of our result.  Dynamical
instabilities caused by planetary mass loss would result in collisions
that leave behind systems with large mutual inclinations and high
eccentricities inconsistent with the small mutual inclinations and low
eccentricities inferred for Kepler-discovered system of multiple planets
\citep[e.g.,][]{Wu2013,Fabrycky2014,VanEylen2015,Mills2019,He2020}.

\begin{figure*}
\plotone{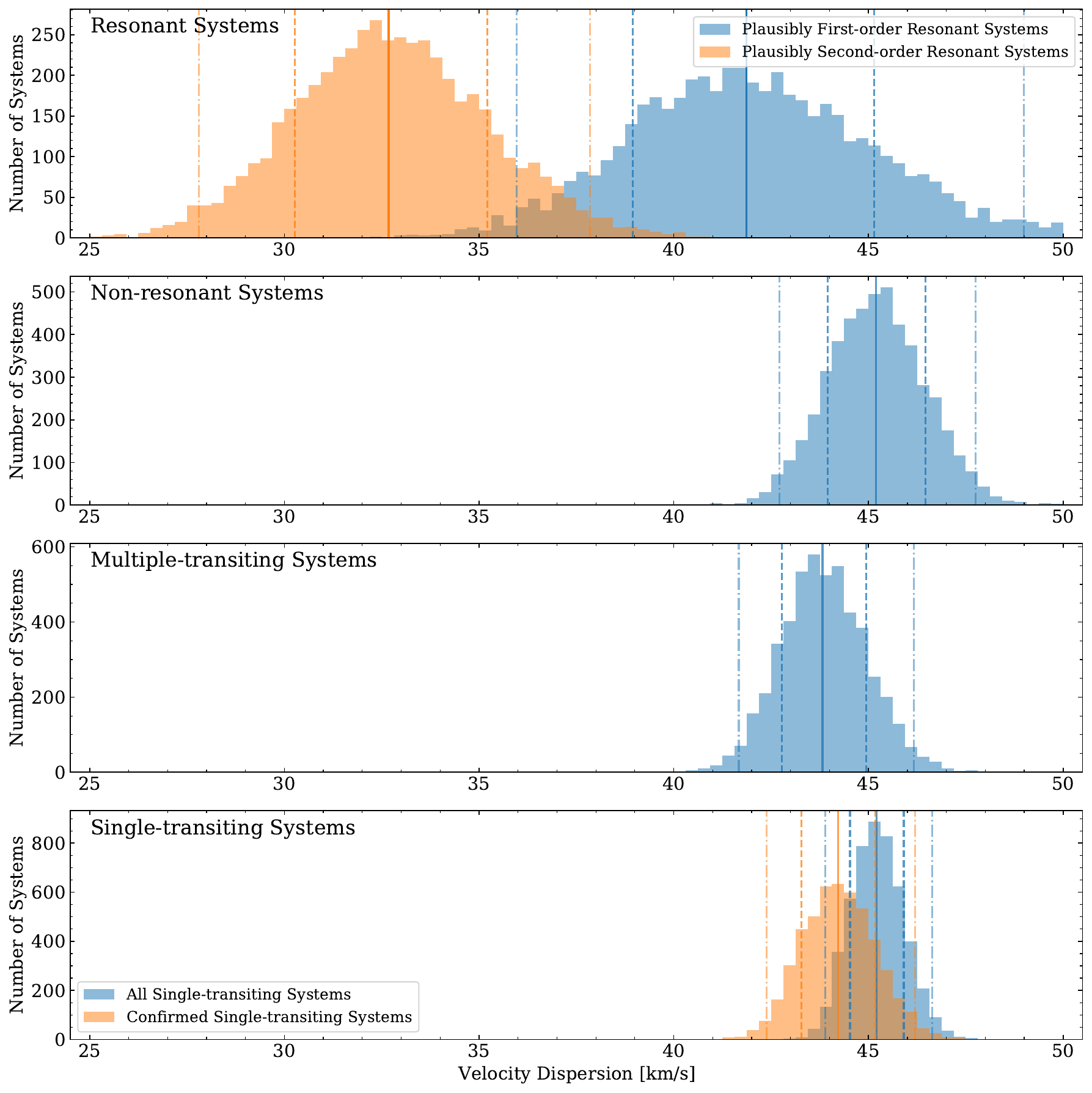}
\caption{Distributions of Galactic velocity dispersions for the
indicated populations.  We indicate median values as solid vertical
lines, 1-$\sigma$ ranges between vertical dashed lines, and 2-$\sigma$
ranges between vertical dot-dashed lines.  Top: the populations
of plausibly first-order (blue) and second-order (orange) resonant
multiple-planet systems.  Top-middle: the population of implausibly
resonant multiple-planet systems.  Bottom-middle: the population of
all multiple-planet systems.  Bottom: the populations of confirmed
(orange) and candidate (blue) single-transiting systems.  The plausibly
second-order resonant system population has a significantly colder
Galactic velocity dispersion than the single-transiting system population.
The population of single-transiting systems has a Galactic velocity
dispersion consistent with the populations of implausibly resonant
multiple-planet systems and all multiple-planet systems.\label{figure10}}
\end{figure*}

Dynamical instabilities do represent a possible
solution to the so called ``Kepler dichotomy'' though
\citep[e.g.,][]{Johansen2012,Izidoro2017,Izidoro2021}.  In this scenario,
single-planet systems may have initially been multiple-planet, possibly
resonant systems that experienced a dynamical instability that excited
eccentricities and inclinations.  If this scenario is accurate, then
single-planet systems should be older than multiple-planet systems.
To evaluate this proposed explanation for the Kepler dichotomy, we use
the \citet{Thompson2018} Kepler DR25 KOI list and compare the Galactic
velocity dispersions of the populations of single-transiting systems,
plausibly first-order/second-order resonant multiple-planet systems,
implausibly resonant multiple-planet systems, and all multiple-planet
systems.  We show the results of these calculations in Figure
\ref{figure10}.  The population of single-transiting systems has a
Galactic velocity dispersion significantly warmer than the population of
plausibly second-order resonant multiple-planet systems.  The Galactic
velocity dispersions of single-transiting, first-order resonant,
implausibly resonant multiple-planet, and all multiple-planet systems are
all consistent.  Our results contrast with LAMOST-based inferences that
found Galactic velocity dispersion decreases with planet multiplicity
\citep{Chen2021a,Chen2021b}.  We find no systematic differences between
single-transiting and multiple-transiting systems in lithium equivalent
width, lithium abundance, $\log{R'_{\text{HK}}}$, or $S_\text{HK}$.
In accord with the idea that single-transiting systems are the outcomes
of dynamical instabilities, these analyses suggests that plausibly
second-order resonant systems are younger than single-transiting systems.
Plausibly second-order resonant systems aside, our analyses provide no
evidence that single-transiting systems are older than multiple-transiting
systems in general.

\subsection{Interactions with Drifting Planetesimals as a Possible
Explanation}

Interactions with a residual disk of planetesimals are a mechanism that
can explain the observations that planet pairs tend to favor period ratios
just wide of resonances and tend to avoid period ratios just interior to
resonances.  \citet{Chatterjee2015} and \citet{Ghosh2023} have simulated
the interaction of resonant planet pairs with co-located planetesimal
disks remaining after the dissipation of gaseous protoplanetary disks.
Those analyses have shown that planets interacting with a total mass
of planetesimals at least 1\% of the total planetary mass can produce
departures from resonances in less than 1 Myr similar in magnitude to
those observed by Kepler.  While that timescale is too short to explain
our results, the arrival of a net planetesimal mass 1\% of the total
planetary mass over a 1 Gyr interval could in principle reproduce both
the Kepler period-ratio distribution and our observation that plausibly
second-order resonant systems have a colder Galactic velocity dispersion
than implausibly resonant systems.

A delayed interaction with an exterior disk of residual planetesimals is
a key ingredient for explaining the architecture of the solar system in
the Nice model \citep[e.g.,][]{Gomes2005,Morbidelli2005,Tsiganis2005}.
It has also been suggested that it would take a few hundred million
years for the terrestrial planets to clear the inner solar system of
small bodies and a few billion years for the gas \& ice giants to clear
the outer solar system of small bodies \citep[e.g.,][]{Goldreich2004}.
Simulations of interactions between Jupiter, Saturn, and a residual
planetesimal disk have shown that the timescale for the initiation of
planetesimal-driven migration is sensitive to the location of the inner
edge of the residual planetesimal disk, and can take more than 1 Gyr
\citep[e.g.,][]{Gomes2005,Thommes2008}.

While \citet{Chatterjee2015} and \citet{Ghosh2023} considered
instantaneous interactions with co-located planetesimals, extended
interactions with migrating planetesimals could also take place.
Planetesimals can migrate long distances over billions of years
due to a process called the Yarkovsky effect in which a rotating
planetesimal heated on its day side by its host star will experience a
non-gravitational acceleration as its hotter day side radiates more than
its cooler night side (prograde rotators migrate outwards and retrograde
rotators migrate inwards).  The net force experienced by a planetesimal
from the Yarkovsky effect has complex dependencies on its thermal
parameters, albedo, emissivity, size, rotation rate, and obliquity.
Yarkovsky effect-driven accelerations can cause significant semimajor axis
evolution for planetesimals with sizes from meters to tens of kilometers.

To evaluate the potential of the Yarkovsky effect to deliver planetesimal
mass into the vicinities of Kepler-observed multiple-planet system over
Gyr timescales, we first roughly estimate the total mass in planetesimals
necessary to deliver via the Yarkovsky effect planetesimals amounting to
about 1\% of the typical total planetary mass in our sample of plausibly
second-order resonant systems.  Assuming the mass--radius relation from
\citet{Lissauer2011}, the typical total planet mass in our sample of
plausibly second-order resonant systems is $M_{\text{p,tot}} \approx
10~M_{\oplus}$.  Assuming planetesimal properties like those in the solar
system, the Yarkovsky effect most strongly affects planetesimals with
sizes $R$ in the range $1~\text{km} \lesssim R \lesssim 10~\text{km}$.  To
roughly calculate the total disk mass necessary to provide $M_{\text{tot}}
\approx 0.1~M_{\oplus}$ of migrating 1 or 10 km planetesimals, we assume
an equal number of prograde/retrograde rotators and first assume the
planetesimal size distribution produced by detailed simulations of
the streaming instability $dN/dR \propto R^{-2.8}$ \citep{Simon2016}.
If objects with $R = (1,10)$ km dominate the migrating planetesimal
mass budget, then we find that a total disk mass $M_{\text{tot}} \approx
(32,20)~M_{\oplus}$ would be necessary to supply sufficient planetesimal
mass to affect period ratios.  We next assume the planetesimal size
distribution appropriate for steady-state collisional evolution $dN/dR
\propto R^{-3.5}$ \citep{Wyatt2008}.  If objects with $R = (1,10)$
km dominate the migrating planetesimal mass budget, then we find that
a total disk mass $M_{\text{tot}} \approx (3,9)~M_{\oplus}$ would be
necessary to supply sufficient planetesimal mass to affect period ratios.
Because we assumed that only planetesimals with a single size migrate,
these are upper limits on the total disk mass.

We use the simplified model of the Yarkovsky effect presented
in \citet{Veras2019} and implemented in \texttt{REBOUNDx 3.1.0}
\citep{Tamayo2020} to infer an upper limit on the amount of migration
experienced by planetesimals as a function of $R$.  We first assume
that the semimajor axes of planetesimals migrating inwards shrink until
they reach the exterior 2:1 mean-motion resonance with the outermost
planet in an initially second-order resonant system.  Upon reaching that
location, an inwardly migrating planetesimal will experience eccentricity
excitation leading to orbit crossing and eventual collision with a planet
in the system.  In our sample of plausibly second-order resonant systems,
the typical semimajor axis of the outermost observed planet is $a =
0.15$ AU.  The corresponding exterior 2:1 mean-motion resonance would
be located at $a_{1} = 0.24$ AU.  We assume a planetesimal density
$\rho_{\text{plan}} = 2.7$ kg m$^{-3}$ and a Bond albedo  $A = 0.017$.
The median stellar mass in our sample of plausibly second-order resonant
systems $\overline{M}_{\ast} = 0.92~M_{\odot}$ corresponding to a
luminosity $L_{\ast} \approx 0.72~L_{\odot}$.  We evolve the orbits of
planetesimals with 1 km and 10 km for 0.1 Myr at an array of semimajor
axes values to estimate the instantaneous $da/dt$ at many $a$.  At time
$t = 0$ we place a planetesimal at the location of the exterior 2:1
mean-motion resonance for our median system and integrate its orbit
backward in time for 1 Gyr.  We find that the migration as a result
of the Yarkovsky effect can move (1,10) km planetesimals to $a_{1} =
0.23$ AU from $a_{0} = (0.45,0.26)$ AU in 500 Myr and from $a_{0} =
(0.62,0.29)$ in 1 Gyr.  Our inferred total planetesimal disk masses and
migration rates imply surface densities $\Sigma_{\text{plan}}$ in the
range $10^2~\text{g cm}^{-3} \lesssim \Sigma_{\text{plan}} \lesssim
10^3~\text{g cm}^{-3}$ consistent with the solid surface density
inferred for the minimum-mass solar nebula $10^1~\text{g cm}^{-3}
\lesssim \Sigma_{\text{MMSN,solid}} \lesssim 10^2~\text{g cm}^{-3}$
\citep[e.g.,][]{Weidenschilling1977,Asplund2009}.

While the total amount of solid mass and the solid surface density
required for Yarkovsky effect-driven planetesimal migration are
plausible, maintaining enough mass in (1,10) km planetesimals for 1
Gyr could be challenging.  Planetesimal disk mass inferences based on
observable dust properties suggest that planetesimal disks can contain
$M_{\text{tot}} \sim 100~M_{\oplus}$ shortly after protoplanetary disk
dissipation \citep[e.g.,][]{Wyatt2008}.  Those masses are expected to
drop below $M_{\text{tot}} \sim 1~M_{\oplus}$ after about 10 Myr due
to collisional grinding.  At the same time, apparent episodes of late
accretion in the solar system provides evidence for planetesimals disks
lasting for a few hundred Myr.  Indeed, a residual disk of planetesimals
with $10~M_{\oplus} \lesssim M_{\text{tot}} \lesssim 100~M_{\oplus}$
in the outer solar system is thought to be necessary to explain the
orbital properties of the gas and ice giants \citep[e.g.,][]{Hahn1999}.
Likewise, a residual disk of planetesimals in the inner solar system with
$M_{\text{tot}} \sim 1~M_{\oplus}$ is thought to be necessary to damp
the eccentricities and inclinations of the terrestrial planets after
the giant impact phase of planet formation a few tens of Myr after the
formation of calcium-aluminum rich inclusions at the dawn of the solar
system \citep[e.g.,][]{Schlichting2012}.  While more detailed simulations
will be necessary to confirm or reject the scenario outlined above,
these solar system-based inferences indicate that our scenario is at
least a plausible explanation for our observations.

\section{Conclusion}

We find that while plausibly second-order mean-motion resonant
multiple-planet systems discovered by the Kepler Space Telescope are not
young in an absolute sense, they do have a colder Galactic velocity
dispersion and are therefore younger than both Kepler-discovered
implausibly resonant multiple-planet systems and single-transiting
systems.  The same is true for first-order mean-motion resonant systems,
but only when tidal dissipation inside a system's innermost planet is
likely important for the system's secular evolution.  Our observations are
inconsistent with any proposed physical mechanisms that drive initially
resonant systems away from resonance in a gaseous protoplanetary disk,
during its dissipation, or that take place in the first 100 Myr of a
system's evolution.

We confirm that the age offset between plausibly second-order resonant and
implausibly resonant systems persists at high statistical significance
even when tidal dissipation inside the innermost planet is likely too
weak to act as an angular momentum sink.  The implication is either
that tides are more efficient than expected (perhaps due to obliquity
tides) or that an additional non-tidal secular process that acts
over hundreds of Myr to one Gyr is responsible for moving initially
second-order resonant systems out of resonance while leaving them
with small eccentricities and mutual inclinations.  We propose that
interactions with km-sized planetesimals migrating inward due to the
Yarkovsky effect from a residual planetesimal belt at $a \sim 1$ AU with
total mass $1~M_{\oplus} \lesssim M_{\text{tot}} \lesssim 10~M_{\oplus}$
are a plausible explanation consistent with our observations and worthy
of more careful future study.  Based on our observation that plausibly
second-order resonant systems are younger than both implausibly resonant
multiple-planet and single-transiting systems, we predict that plausibly
second-order resonant systems should be frequently found orbiting young
stars with ages between about 10 Myr and 100 Myr.  The same should be
true for first-order resonant systems with an innermost planet likely
affected by tidal dissipation.  Both plausibly first- and second-order
resonant systems should then become less common as systems age.

\section*{Acknowledgments}

This material is based upon work supported by the National Science
Foundation under grant number 2009415.  We thank Jason Steffen
for helpful comments and for providing their catalog of Kepler
multiple-planet systems in advance of publication.  We also thank
the anonymous referee for an insightful suggestion that significantly
improved this article.  This research has made use of the NASA Exoplanet
Archive \citep{Akeson2013}, which is operated by the California Institute
of Technology, under contract with the National Aeronautics and Space
Administration under the Exoplanet Exploration Program.  This research
has made use of the SIMBAD database, operated at CDS, Strasbourg, France
\citep{Wenger2000}.  This work has made use of data from the European
Space Agency (ESA) mission Gaia (\url{https://www.cosmos.esa.int/gaia}),
processed by the Gaia Data Processing and Analysis Consortium (DPAC,
\url{https://www.cosmos.esa.int/web/gaia/dpac/consortium}).  Funding for
the DPAC has been provided by national institutions, in particular the
institutions participating in the Gaia Multilateral Agreement.  This
research has made use of the VizieR catalog access tool, CDS, Strasbourg,
France (\href{http://doi.org/10.26093/cds/vizier}{10.26093/cds/vizier}).
The original description of the VizieR service was published in
\citet{Ochsenbein2000}.  Funding for SDSS-III has been provided by the
Alfred P. Sloan Foundation, the Participating Institutions, the National
Science Foundation, and the U.S. Department of Energy Office of Science.
The SDSS-III web site is \url{http://www.sdss3.org/}.  SDSS-III is
managed by the Astrophysical Research Consortium for the Participating
Institutions of the SDSS-III Collaboration including the University
of Arizona, the Brazilian Participation Group, Brookhaven National
Laboratory, Carnegie Mellon University, University of Florida, the French
Participation Group, the German Participation Group, Harvard University,
the Instituto de Astrof\'isica de Canarias, the Michigan State/Notre
Dame/JINA Participation Group, Johns Hopkins University, Lawrence
Berkeley National Laboratory, Max Planck Institute for Astrophysics,
Max Planck Institute for Extraterrestrial Physics, New Mexico State
University, New York University, Ohio State University, Pennsylvania
State University, University of Portsmouth, Princeton University, the
Spanish Participation Group, University of Tokyo, University of Utah,
Vanderbilt University, University of Virginia, University of Washington,
and Yale University.  Funding for the Sloan Digital Sky Survey IV has
been provided by the Alfred P. Sloan Foundation, the U.S.  Department
of Energy Office of Science, and the Participating Institutions.
SDSS-IV acknowledges support and resources from the Center for High
Performance Computing at the University of Utah.  The SDSS website is
\url{www.sdss.org}.  SDSS-IV is managed by the Astrophysical Research
Consortium for the Participating Institutions of the SDSS Collaboration
including the Brazilian Participation Group, the Carnegie Institution
for Science, Carnegie Mellon University, Center for Astrophysics |
Harvard \& Smithsonian, the Chilean Participation Group, the French
Participation Group, Instituto de Astrof\'isica de Canarias, The Johns
Hopkins University, Kavli Institute for the Physics and Mathematics of
the Universe (IPMU) / University of Tokyo, the Korean Participation Group,
Lawrence Berkeley National Laboratory, Leibniz Institut f\"ur Astrophysik
Potsdam (AIP),  Max-Planck-Institut f\"ur Astronomie (MPIA Heidelberg),
Max-Planck-Institut f\"ur Astrophysik (MPA Garching), Max-Planck-Institut
f\"ur Extraterrestrische Physik (MPE), National Astronomical Observatories
of China, New Mexico State University, New York University, University of
Notre Dame, Observat\'ario Nacional / MCTI, The Ohio State University,
Pennsylvania State University, Shanghai Astronomical Observatory,
United Kingdom Participation Group, Universidad Nacional Aut\'onoma
de M\'exico, University of Arizona, University of Colorado Boulder,
University of Oxford, University of Portsmouth, University of Utah,
University of Virginia, University of Washington, University of Wisconsin,
Vanderbilt University, and Yale University.  This research made use of
Astropy\footnote{\url{http://www.astropy.org}}, a community-developed
core Python package for Astronomy \citep{astropy2013,astropy2018}.
This research has made use of NASA's Astrophysics Data System.


\vspace{5mm}
\facilities{ADS, Exoplanet Archive, Gaia, Kepler, LAMOST, Sloan}

\software{\texttt{Astropy} \citep{astropy2013,astropy2018},
          \texttt{galpy} \citep{Bovy2015},
          \texttt{matplotlib} \citep{matplotlib},
          \texttt{pandas} \citep{pandas},
          \texttt{pyia} \citep{PriceWhelan2018},
          \texttt{REBOUNDx} \citep{Tamayo2020}}

\clearpage
\bibliography{ms}{}
\bibliographystyle{aasjournal}

\end{document}